\documentclass[conference]{IEEEtran}
\IEEEoverridecommandlockouts

\usepackage{times}

\usepackage{makecell}
 
\usepackage{graphicx}
\usepackage{amsmath}
 
\usepackage{algorithm}
\usepackage{algorithmic}
 
\usepackage{graphicx}
\usepackage{subfigure}

\usepackage{times}
\usepackage{epsfig}
\usepackage{graphicx}
\usepackage{amsmath}
\usepackage{amssymb}

\usepackage{bm}
\usepackage{graphicx}
\usepackage{times}

\usepackage{algorithm}
 
\usepackage{amsmath}
 
\usepackage[english]{babel}
\usepackage{graphicx}
\usepackage{epsfig}
\usepackage{epstopdf}
 
\usepackage{amsmath}
 
\usepackage[english]{babel}
\usepackage{graphicx}
\usepackage{graphicx}
\usepackage{epsfig}
\usepackage{epstopdf}
 
\usepackage{amssymb}
\usepackage{subfigure}
\usepackage{color}
\newtheorem{theorem}{Theorem}

\newtheorem{fact}{Fact}
\newtheorem{proof}{Proof}
\newtheorem{example}{Example}

\def\BibTeX{{\rm B\kern-.05em{\sc i\kern-.025em b}\kern-.08em
		T\kern-.1667em\lower.7ex\hbox{E}\kern-.125emX}}
\begin{document}

\title{The Scalability for Parallel Machine Learning Training Algorithm: Dataset Matters}
 
\author{\IEEEauthorblockN{Daning Cheng \IEEEauthorrefmark{1}\IEEEauthorrefmark{2},
		Hanping Zhang\IEEEauthorrefmark{3}\IEEEauthorrefmark{4},
		Fen Xia\IEEEauthorrefmark{3}, 
		Shigang Li\IEEEauthorrefmark{5} and
		Yunquan Zhang\IEEEauthorrefmark{1}}
	\IEEEauthorblockA{\IEEEauthorrefmark{2}University of Chinese Academy of Sciences, Beijing, China
	}
	\IEEEauthorblockA{\IEEEauthorrefmark{1}SKL of Computer Architecture, Institute of Computing Technology, CAS, China\\
		Email:  \{chengdaning,zyq\}@ict.ac.cn}

	\IEEEauthorblockA{\IEEEauthorrefmark{5}Department of Computer Science,
		ETH Zurich,Switzerland \\
		Email: shigangli.cs@gmail.com}
	\IEEEauthorblockA{\IEEEauthorrefmark{4}University at Buffalo, State University of New York}
	\IEEEauthorblockA{\IEEEauthorrefmark{3}Wisdom Uranium technology Co.Ltd, Beijing\\
		{\{ Xiafen, Zhanghanping \}@ebrain.ai}}
}

\maketitle
\begin{abstract}
To gain a better performance, many researchers put more computing resource into an application. However, in the AI area,  there is still a lack of a successful large-scale machine learning training application: The scalability and performance reproducibility of parallel machine learning training algorithm, i.e. stochastic optimization algorithms,  are limited and there are  few pieces of research focusing on why these indexes are limited.

In this paper, we propose that the sample difference in dataset plays a more prominent role in parallel machine learning algorithm scalability. Dataset characters can measure sample difference. These characters include the variance of the sample in a dataset, sparsity, sample diversity and similarity in sampling sequence. 

To match our proposal, we choose four kinds of parallel machine learning training algorithms as our research objects: (1) Asynchronous parallel SGD algorithm (Hogwild! algorithm) (2) Parallel model average SGD algorithm (Mini-batch SGD algorithm) (3) Decenterilization optimization algorithm, (4) Dual Coordinate Optimization (DADM algorithm).  These algorithms cover different types of machine learning optimization algorithms.

We present the analysis of their convergence proof and design experiments. Our results show that the characters datasets decide the scalability of the machine learning algorithm. What is more, there is an upper bound of parallel scalability for machine learning algorithms.
\end{abstract}

\begin{IEEEkeywords}
	Parallel algorithm, Scalability, Dataset,  Stochastic optimization method
\end{IEEEkeywords}

\section{Introduction}
Training a machine learning model is an exhausting job. Training a machine model often uses stochastic optimization methods, like stochastic gradient descent,  stochastic dual coordinate ascent method.  With the development of parallel computing methods, to reduce training time, parallel and distribution optimization methods are proposed. Nowadays, machine learning frameworks, which use these distribution optimization methods, are widely used in AI and machine learning industry like MXNet, Tensorflow.

However, the scalability for those machine learning frame and algorithm is limited:

1. Although researchers offered state of the art large-scale machine learning training applications, those applications do not run machine learning training processes on the large-scale parallel system: Some jobs apply machine learning as a  part of a large-scale parallel system. For example, in work by H.Wang et al\cite{WangDeePMD}, the author trains a DNN on a desktop computer and put this DNN into LAMMPS, which is a software which runs on HPC. 

2. Although large-scale computing device, like GPU,  is well-used in machine learning training process, yet that large-scale computing is limited in traditional math kernels. Some works try their best to optimize math kernel, like matrix multiplication(GEMM kernel). For example, In the work by S.Chetlur et al. \cite{Chetlur2014cuDNN}, the authors designed the methods which GPU uses to compute convolution operation. 

3.  For general cases, with the more parallel computing resource throwing into those machine learning frame, it is evident that the effect of those frameworks does not improve too much. Some works claim that current distribution machine learning frameworks can only contain less than 100 nodes or GPU\cite{AnilLarge}. In work by R.Anil et al\cite{AnilLarge}, they claim that "it can be very difficult to scale effectively much beyond a hundred GPU workers in realistic setups." For mini-batch SGD, few works try to use the setting whose batch-size is larger than 32K\cite{you2017imagenet}.

4. Some works focus on training specific machine learning models on a particular dataset. For example,  some researchers use a specific DNN training Imagenet dataset, but their work cannot be pushed into other machine learning models and datasets \cite{you2017imagenet} \cite{you2017scaling}\cite{qixiang}. In work by Thorsten Kurth et al\cite{qixiang}, they trained a specific DNN on a specific dataset. However, they do not show their scalability performance can be pushed into other DNN and datasets. 

5. Some works use parallel pipeline methods which break goal function, like DNN, into several pieces, and those pieces are computed in a pipeline system\cite{qixiang}. However, there is no mathematical basis for this parallel optimization method. What is more, recent work by Igor Colin et al\cite{NIPS2019_9402} shows that in mathematical,  only non-smooth problems may benefit from pipeline parallelization. However, even in a complex machine learning model training process, like a DNN training process,  non-smooth goal function is still rare, because to gain gradient easily, most of non-smooth part of goal functions are replaced by smooth functions. For example, the step function is replaced by sigmod function in DNN.

So, current parallel machine learning works are unsatisfied: (1) The improvement of distributed parallel machine learning is small with a large parallel computing resource. In some cases, the influence of using a  large parallel computing resource can be harmful. (2) Many works are lack of replicability. The scalability performance for a specific machine learning model on the specific dataset cannot be pushed into other models or datasets.

Thus, we proposed the question that: are the current state of the art parallel optimization methods able to run on super large-scale parallel computing environment? Besides the size of the dataset and engineering implements, are there any other factors which play critical roles on the scalability for parallel machine learning training algorithm?

To solve the above questions, in this paper, we choose four different kinds of state of the art parallel optimization methods as our benchmarks: (1) Asynchronous parallel SGD algorithm, i.e., ASGD (Hogwild! algorithm)\cite{Niu2011HOGWILD}, (2) Parallel model average SGD algorithm (Mini-batch SGD algorithm\footnote{In mini-batch SGD, parallel technology is used in parallel computing mini-batch gradient: one sample's gradient computing process uses one worker/thread. In other word, the maximum degree of parallel is the size of batch-size. Full algorithm description can be seen in the appendix in https://arxiv.org/abs/1910.11510 or https://github.com/tomcheng0618/dataset})\cite{Duchi2016Intro}. (3) Decentralization optimization algorithm (ECD-PSGD)\cite{tang2018decentralization} and (4) Dual Coordinate Optimization (Distributed Alternating Dual Maximization algorithm, abbr. DADM)\cite{Zheng2017A}. 

After examining the convergence analysis of the above algorithms, we find that when the machine learning model is fixed, the sample difference plays a vital role in the scalability. Some characters of the dataset can describe sample differences. Those characters include(1) the variance of sample feature in a dataset. (2) the sparsity of the sample in a dataset. (3) the diversity of the sample in a dataset. (4) the similarity of successive sampling samples.

What is more, we also find that for most of the algorithms, the gain growth is minor with the increasing of using the parallel resource in mathematics. The above fact shows that for most of the training algorithms, there is the scalability upper bound.

To prove our analysis, we conduct experiments. We design and choose different datasets that share different characters. Our experiment results match the convergence proof analysis. 

Our contribution is summarized as follows:

1. We examined four different state-of-the-art parallel optimization methods. In view of convergence analysis, we show that the sample difference, which can be described by the characters of the dataset, plays crucial roles in parallel machine learning algorithms scalability. The dataset characters at least include (1) the variance of sample feature in a dataset. (2) the sparsity of the sample in a dataset. (3) The diversity of the sample in a dataset. (4) the similarity of successive sampling samples.

2. Different datasets suit different optimization methods.

3. The scalability of the stochastic optimization algorithm has its upper bound, which is decided by the dataset.

4. We design experiments to prove the importance of the dataset on parallel machine learning algorithms scalability.  We also show the upper bound of algorithm scalability on experiment datasets.

Our analysis and experimental results answer the following problem: 

1. The current parallel machine learning algorithms cannot make full use of a large-scale parallel computing environment, like a supercomputer. These large parallel computing environments' parallel degrees are much higher than the upper bound of algorithm scalability.

2. One scalability performance of an algorithm on a specific dataset cannot be pushed into other datasets. 

3. To improve scalability,  random sort for datasets is necessary.

\section{Related works}
With the development of parallel computing and optimization methods, many parallel optimization methods are designed to make the machine learning training process fast. The most widely used methods are different parallel SGD algorithm, and the newest state-of-the-art methods include decentralization algorithm and dual optimization algorithms.

\subsection{Parallel SGD algorithm}
SGD can be dated back to the early work of Robbins and Monro \cite{Robbins1951A,Ermoliev1969On,Nemirovski2009Robust,Polyak1992Acceleration,Bottou2010Large}. Recent years, combining with the GPU and clusters \cite{TensorFlowTensorFlow,Jia2014Caffe}, parallelized SGD became the most powerful weapon solving machine learning problems \cite{Bertsekas2003Parallel,Dekel2012Optimal,Duchi2010Distributed,Langford2009Slow}.Parallel SGD can be roughly classified into two categories - Asynchronous parallel  SGD, like Hogwild! \cite{Feng2011HOGWILD} and Model Average Parallel SGD like mini-batch SGD and Simul Parallel SGD\cite{Zinkevich2010Parallelized}.

The goal results for the Asynchronous Parallel SGD algorithm and sequential SGD are the same in fixed iterations.  Model Average Parallel SGD algorithms give us the answer about how to calculate a better output in a fixed number of iterations. 

Decentralized parallel stochastic gradient descent\cite{yuan2016on} is one kind of the decentralization algorithm. Decentralized parallel stochastic gradient descent requires each node to exchange its own stochastic
gradient and update the parameter using the information it receives\cite{tang2018decentralization}.

\subsection{Dual Coordinate Ascent Optimization}

Stochastic dual coordinate ascent method(SDCA) \cite{Shalevshwartz2014Accelerated} \cite{Shalevshwartz2013Stochastic} is one of the most important optimization method.  Its data parallelism algorithms are hot topic in optimization algorithm area\cite{Jaggi2014Communication}\cite{Ma2015Adding}.  DADM\cite{Zheng2017A}, DisDCA \cite{yang2013trading}, CoCoA+\cite{ma2017distributed} are state of the art distribution parallel Dual Coordinate Optimization.

\section{Background}
\subsection{Problem Setting}
For machine learning, an optimization method is used to solve the following minimum problem:
\begin{equation*}
min \hat{f}(x)=\mathbb{E}_{\Xi } F(x; \Xi)
\end{equation*} 
where $\Xi$ is a random variable that satisfies a certain distribution. In most cases, the distribution of $\Xi$ is unknown or cannot be presented as a formula form. It is common that we use the frequency histogram to replace $\Xi$'s PDF. Above formula is written as:
\begin{equation}
min\hat{f}(x)=\mathbb{E}_{\Xi } F(x;\Xi)\approx f(x)=\frac{1}{n}\sum_{i=1}^{n}F(x;\xi_i) \label{appro}
\end{equation}  
where $\xi_i$ is the sample which sampled from $\Xi$. The collection of $\{\xi_1,\xi_2,...,\xi_n\}$ is the dataset.  And $x^*=\mathop{argmin}f(x)$.

For regularized risk minimization, $F(x;\xi_i)$ is usually presented as following formula\cite{Zinkevich2010Parallelized}:
\begin{equation*}
F(x;\xi_i)=L(\xi_i,x)+\lambda\psi(x)%\frac{\lambda}{2}\| x \|^2
\end{equation*}
$L(\xi_i,x)$ is the loss function like hinge loss for SVM model and logloss for LR model, $\psi(x)$ is regulation function. Usually, $\psi(x)=\frac{1}{2}\| x \|^2$, i.e.

\begin{equation}
F(x;\xi_i)=L(\xi_i,x)+\frac{\lambda}{2}\| x \|^2
\label{reg_goal}
\end{equation}

\subsection{Sample Uniformly Distribution Assumption}
To make analysis simple and clear, we have to assume that the distributions of the sample's feature value and non-zero feature's position are uniform. As we can see,   most of the datasets satisfy this assumption.  

We have to use this assumption because, without it, common definitions may blur the boundary between different types of dataset, as the following example.

\begin{example}
	Considering the dataset: \{(1,0,0,...,0), (2,0,0,...,0), (3,0,0,...,0) ... (dataset\_size,0,0,...,0)\}, under the common definitions, this dataset is sparse dataset. However, after delete unused features, above dataset is a dense dataset.
\end{example}

\section{The index to measure sample difference} 
\subsection{the local similarity of consecutive samples in the sampling sequence, i.e.,  $LS_{\mathcal{A}}(\mathcal{D},\mathcal{S})$,  for different algorithm}

We find that similarity of consecutive samples in the sampling sequence is important, because in online learning applications which often use SGD as their optimization method, rearranging samples always lead to a better scalability performance. In an online learning application, the samples in the sample sequence are often similar to their neighborhood samples. For example, the online sample from advertisement click is similar to its neighborhood, because user interest cannot be changed drastically. 

To make our presentation clearly, we have to define the local similarity, i.e. $LS_{\mathcal{A}}(\mathcal{D},\mathcal{S})$, for algorithm $\mathcal{A}$ on dataset $\mathcal{D}$ using sampling sequence $\mathcal{S}$.

Before we define the local similarity of consecutive samples in the sampling sequence, i.e. $LS_{\mathcal{A}}(\mathcal{D},\mathcal{S})$, we have to define the value of $C\_{sim}$.

For a sampling sequence $\xi_1,\xi_2,....,\xi_n$ and a range $range$,  $C_{sim}$ is defined as
\begin{equation}
C\_sim_{range}=\frac{1}{n}\sum_{i=1}^{n}\frac{\sum_{j=1}^{range}\left\|\xi_i-\xi_{(i+j)\%length} \right\|_0}{range}
\end{equation}
where $length$ is the sequence length.

%In our following analysis, we would conclude that break similarity would gain better scalability. $C\_sim$ is the parameter that measures similarity. 

For a sample collection $\{\xi_1,\xi_2,..,\xi_n\}$, their different sampling orders have different $C\_sim$.
\begin{example}
	For dataset (0,0,0), (0,0,1), (0,1,1), (0,1,0), (1,1,0), (1,0,0), the samples have 2 different $C\_sim_2$ sequence:
	
	1. Sequence with $C\_sim_2$=0.5:  (0,0,0), (0,0,1), (0,1,1), (0,1,0), (1,1,0), (1,0,0)

	2. Sequence with $C\_sim_2$=1: (0,0,0), (1,1,0), (0,0,1), (1,0,0), (0,1,0),(0,1,1)
\end{example}

Based on the $C\_sim_{range}$, we can give the definition of $LS_{\mathcal{A}}(\mathcal{D},\mathcal{S})$.

When $\mathcal{A}$ is an asynchronous SGD like Hogwild!,  the $\mathcal{S}$ is   $\xi_1,\xi_2,...,\xi_t$, $\xi_i \in \mathcal{D}$, and the jag between when a gradient is computed and when it is used is always less than or equal to $\tau_{max}$. Then, $LS_{\mathcal{A}}(\mathcal{D},\mathcal{S}) = C\_sim_{\tau_{max}}$ on $\mathcal{S}$.

When $\mathcal{A}$ is a synchronous algorithm, like DADM, mini-batch SGD and ECD-PSGD, the $\mathcal{S}$ is $[ \xi_1,\xi_2,...,\xi_{batch\_size}]$, $[ \xi_{batch\_size+1}$ ,$\xi_{batch\_size+2}$,..,$\xi_{2batch\_size}],$$...$, $\xi_i \in \mathcal{D}$, where the samples or gradients in $[\cdot]$ are in one batch. We use following two steps to calculate $LS_{\mathcal{A}}(\mathcal{D},\mathcal{S})$: 1, For the samples in a batch, we find the sequence which consist of these samples. This sequence's C\_sim$_{batch\_size}$ is larger than it for any sequences which consist of these samples. We name C\_sim$_{batch\_size}$ for this sequence as $C\_sim\_batch$. 2, For whole sampling sequence $\mathcal{S}$, we choose the batch whose $C\_sim\_batch$ is maximum in  $\mathcal{S}$. And $LS_{\mathcal{A}}(\mathcal{D},\mathcal{S})$ is this batch's $C\_sim\_batch$.

%the sample sequence $\mathcal{S}$ we discuss is the sequence which consist of one of sequence 's mini-batch, can build by an sample batch  and we pick the sequence  which can build the maximum $C\_sim_{batch\_size}$ in one sample batch, i.e., $LS_{\mathcal{A}}(\mathcal{D},\mathcal{S}) = \mathop{max} C\_sim_{batch\_size}$

%\begin{example}
%     In mini-batch SGD which batch\_size is 3, $\mathcal{S}$ is $\{\xi_1,\xi_2,\xi_3\}$,$\{\xi_4,\xi_5,\xi_6\}$,...,  $\{\xi_{3t-2},\xi_{3t-1},\xi_{3t}\}$, where the sample or gradient in $\{\cdot,\cdot,\cdot\}$ is in one batch. Then, the $LS_{\mathcal{A}}(\mathcal{D},\mathcal{S})$  for mini-batch SGD algorithm is the $C\_sim_{batch\_size}$ for $\xi_{3k+1},\xi_{3k+2},\xi_{3k+3}$ and $\xi_{3k+1},\xi_{3k+2},\xi_{3k+3}$ can build  a sequence whose $C\_sim_{batch\_size}$ is the maximum in all batches.
%\end{example}

\subsection{Feature variance and sparsity}
In this paper, we define the variance of feature $k$ as 
\begin{align*}
&feature \mbox{ }mean_k = \frac{1}{n}\sum_{i=1}^{n}\xi_i(k)\\
&feature\mbox{ }variance_k =  \frac{1}{n}\sum_{i=1}^{n}(\xi(k)- feature \mbox{ }mean_k)^2
\end{align*}
where $\xi_i(k)$ is the k-th feature in $\xi_i$.

The sparsity is the rate between the number of zero elements with the size of the sample. The density is $1 - $
sparsity.

It is clear that when the dataset is sparse, the feature variance must be small.

\subsection{Diversity} The diversity is the number of different kinds of samples in the dataset. We notice that the size of the dataset may be large, but the dataset is the replication of several samples.

Diversity cannot be present by variance and sparsity. Thus, it is necessary to use this index to describe the sample difference.  In the following example, we will show that low variance, low-density datasets still can have high diversity.

\begin{example} Low density dataset whose sample size is large and diversity is high: {(1,0,0,...,0), (0,1,0,...,0), ..., (0,0,0,...,1)}.
\end{example}

\begin{example} The diversity of low variance dataset {(0.01),(0.02),(0.03),...,(0.99),(1)} is higher than the diversity of high variance dataset {(100),(-100),(100),(-100),...,(100),(-100)}.
\end{example}

%\subsection{}

\section{The Upper Bound of Scalability}
\subsection{Perfect Computer Assumption (PCA)}
We want to prove that the reason for limitation of scalability is rooted in parallel stochastic optimization algorithms and datasets. 

So, to avoid the discussion of the code implementation, parallel math kernel implementation, and hardware setting, we assume that the nodes in a cluster have unlimited memory and the bandwidth in the network, and we name this assumption as perfect computer assumption (abbr. PCA). 

The computer with unlimited memory and the bandwidth is named as a perfect computer, and the cluster which consists of perfect computers is named as the perfect computer cluster. 

We can easy to gain the following conclusion: The upper bound of algorithm scalability under the PCA is higher than the upper bound of algorithm scalability on a real computing system. Thus, under this assumption, we can focus on the degree of parallelism, which is offered by the algorithm.

\subsubsection{Iteration and time under the PCA.} We conduct experiments under the PCA also because it is easy to map the number of system/parameters server iteration, $number_{iteration}$, to the real-time. For model average parallel algorithm like mini-batch SGD, ECD-PSGD, DADM, The time for one iteration on single perfect computer is $t_{single}$, the time for a system with $m$ nodes is $t_{single}*number_{iteration}$.  For asynchronous parallel algorithms like Hogwild!, The time for one iteration on a single perfect computer is $t_{single}$, the time for a system with $m$ nodes is $t_{single}/m*number_{iteration}$. 

Thus, in experiments, when it comes to exhibiting the time consumption performance under the PCA, it is enough to exhibit the number of iteration.

\subsection{The Upper Bound of Algorithm Scalability}

\subsubsection{Gain, Cost, and Gain Growth} The cost is the number of iterations for each worker. The gain is the value of goal function at a fixed iteration. 

The gain growth is the value of goal function's difference or the cost difference between using $m$ nodes and $m+1$ nodes at a fixed iteration, and we use them in different cases.

In theory analysis, for ASGD algorithm like Hogwild! and DADM, the gain growth is the difference between the cost. For example,  when we use $m$ workers, each worker trains $n_{local_{m_1}}$ iterations to reach the points of convergence. When we use $m+1$ workers, each worker trains $n_{local_{m_2}}$ iterations to reach the point of convergence. The gain growth is  $n_{local_{m_1}} -n_{local_{m_2}}$.

\begin{example}
	Using the real-sim dataset, eight equal performance workers, and other stable algorithm settings on Hogwild!, the server uses 6242 iterations to reach the point of convergence. In this case, the cost is the number of iterations for each worker: 6242/8 = 781 iterations per worker. Using the real-sim dataset and nine equal performance workers, the server uses 6497 iterations to reach the point of convergence. In this case, the cost is the number of iteration per worker: 6497/9  = 722 iterations per worker. Thus, the gain growth is 781 - 722 = 59 iterations. As we can see from this example, although the server has to train more iterations, yet the number of iterations per worker is decreasing.
\end{example}

In theory analysis, for mini-batch SGD and ECD-PSGD, the gain growth is the value of goal function's difference. For example, log loss decreases between using $m$ nodes and $m+1$ nodes at a fixed iteration.  It is worthy to note that in this definition, the gain growth for ASGD is always negative.

\begin{example} 
	Using the HIGGS dataset, two workers, and other stable algorithm settings on the mini-batch SGD algorithm, at 50 server iteration, the log loss for this model is 4.7525.  Using the HIGGS dataset, three workers, and other stable algorithm settings on mini-batch SGD, at 50 server iteration, the log loss for this model is 4.5871. The gain growth is   $ 4.7525 - 4.5871  = 0.1654$.
\end{example}

\textbf{Note} In fact, above gain growth definitions are two aspects of one phenomenon.  In practice, we can use them both in an algorithm. Apparently,  in one algorithm, the values of gain growth in the above situations are positively related.

However, we use the above two situations in our paper for the following two reasons:

1. The proof methods for algorithms' theory are different. Different algorithms are proven in different aspects. For the synchronous algorithm, it is easy to be proved in the first situation, asynchronous for second for the asynchronous algorithm. So,  we kept them all.

2. In different presentation methods of experiment data, the above two situations present different clarity: first case suits chart (which can be shown easily by the gap between different convergence curves) and second case suits table.

So, based on the above reasons, in theory, the upper bounds of synchronous algorithms are defined as the first case, ASGD for the second case. In experiments, chart results are presented in the first case and table results for the second case.

\subsubsection{The Theory Upper Bound of Algorithm Scalability}  Base on the definition of gain and gain growth, under the PCA, the upper bound of algorithm scalability, $m_{max} $, is to describe the following two situations:

1. Under perfect computer assumption, with the increasing of the number of nodes at the range $[m_{max},\inf]$, the gain growth is positive but close to zero. In this case, the gain growth would not cover the parallel cost on a real computer. This situation suits mini-batch SGD, DADM, and ECD-PSGD.

\begin{example} Using real-sim dataset and other stable algorithm setting on mini-batch SGD,  the gain growths at 15000 iteration is the 0.0011, 0.0006,0.0003,0.0002,0.00018, matching  to the algorithm setting whose number of worker/batch-size is 14,15,16,17,18,19. As we can see from this case, the gain growth is decreasing (to zero). Thus, when the growth cannot cover the parallel cost, the system meets its scalability upper bound.
\end{example}

2. Under perfect computer assumption, with the increasing of the number of nodes at the range $[m_{max},\inf]$, the gain is decreasing, or cost is increasing drastically. This situation suits the algorithms like Hogwild!.

\begin{example} Using the HIGGS dataset and other stable algorithm settings on Hogwild! the algorithm, the gain growth is 14, 4, -7, -39, -72 match to the algorithm setting whose number of workers is 3, 4, 5, 6, 7. As we can see from this case, the gain growth is decreasing. Thus, when gain growth is negative, the system meets its scalability upper bound.
\end{example}

\section{Theory Analysis Conclusion}
In this section, we will show the analysis conclusion for the above algorithms.  Because of the limitation of the paper pages, we offer the proof and detail analysis in Appendix\footnote{https://arxiv.org/abs/1910.11510 or https://github.com/tomcheng0618/dataset}. 
\subsection{Analysis}

After checking the theory, we can gain the following conclusions.

1. Different datasets suit different parallel machine learning training methods. Feature variance, sparsity, and sample diversity can roughly classify datasets into different suitable algorithms. %Besides high diversity datasets suit to DADM,  
For ASGD, like Hogwild!, and mini-batch SGD and ECD-PSGD, we show the following figure \ref{axis}. 
\begin{figure}[htbp]
	\includegraphics[width= \linewidth]{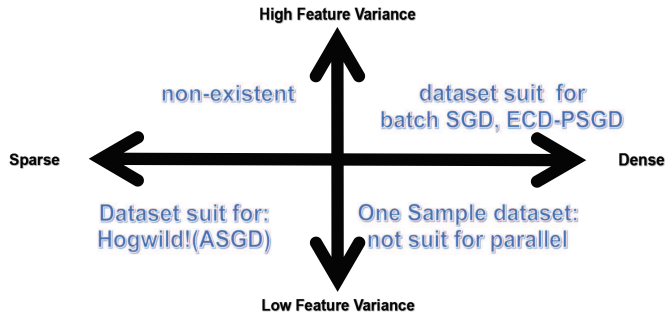}
	\caption{Different datasets suit  different parallel training methods}
	\label{axis}
\end{figure}

2. DADM suits for the dataset whose sample diversity is high.

3. The algorithm scalability performance for the same algorithm can be various depending on $LS_{\mathcal{A}}(\mathcal{D},\mathcal{S})$.

4.  The character of datasets decides the upper bound of algorithm scalability.

Based on our analysis, we can draw the following conclusion clearly:

1. Different datasets suit different parallel stochastic optimization algorithms. 

2. Before training a machine learning model, rearranging samples in the dataset is an excellent choice.

3. No matter which parallel stochastic optimization algorithm is, there always exists an upper bound of scalability. 

4. For the scalability is the character of the dataset and machine learning model, the scalability performance for a particular dataset on a specific machine learning model cannot be pushed into other cases.

\subsection{Current parallel machine learning training algorithm and the traditional problem algorithm in HPC}

When it comes to the scalability of the algorithm, the current machine learning training algorithm and traditional application show a great difference.

For traditional problems, we have the following claims. 

1. The accuracy of output is decided by the number of the grid. Usually, those problems usually can be transformed into linear algebra problems like a stencil, matrix multiplication. The larger problem size we have, the more accurate the results are. 

\begin{example}
	In stencil application, like atmosphere simulation application, the more gird we have, the more accurate the results we can get.
\end{example}

2. The number of the grid, problem size decides how many nodes this application can use, i.e., the number of grid decides the upper bound of theory scalability. The upper bound of theory scalability and real computer environment decide the real upper bound of scalability.

\begin{example}
	When using multi-thread parallel methods to compute a 10*10 vector inner product on a server, which consists of 2 Intel ® Xeon® CPU E5-2680 2.88GHZ, i.e., 24 cores together, we have almost used ten core at most (The upper bound of theory scalability). However, considering the parallel cost, in real situations, we only use one core to solve this problem, the real upper bound of scalability.   
\end{example}

However, for parallel stochastic optimization algorithms, we have the following claims.

1. The accuracy of output is decided by the size of the dataset. Based on the asymptotics in statistics\cite{CamAsymptotics}, in Eq. \ref{appro}, the  larger the training dataset is, the  closer between  $\hat{f}(x)$ and $f(x)$ is.

\begin{example}
	In the real-sim dataset, the log loss on a test dataset of the LR model, which is trained by part of the training dataset, is higher than the LR model, which is trained by the whole training dataset. 
\end{example}

2. The statistical character of the dataset would influence the characters of the machine learning model (goal function in stochastic optimization problem). The characters of the goal function decide the upper bound of scalability. Usually, the size of the dataset may have an influence on the upper bound of scalability. However, the size is dataset is not the decisive element, like the following example.

\begin{example}
	\textbf{One sample dataset:} Considering the dataset which only contains one sample, the size of the dataset can be any number by duplicating this sample. However, the training machine learning model on this dataset cannot be accelerated by any parallel stochastic optimization algorithm. 
\end{example}

\section{Experiment }

In Feature variance and Sparsity Experiment, Diversity Experiment and  $LS_{\mathcal{A}}(\mathcal{D},\mathcal{S})$ Experiment, we use gain to indicate gain growth.

In those experiments, because of the PCA, we will show the convergence curve on the figure whose X-axis is the number of iteration, and the Y-axis is the log loss.

In our figure, the gap can indicate the effect of parallel technology. The upper bound of algorithm scalability has two situations. So, different algorithms have a different index to determine the scalability effect of the parallel algorithm:

For ASGD, i.e., Hogwild!, the effect is better when the gap is smaller, for ASGD's first gain growth definition is always negative. When the gap is small, the number of system iteration to reach a fixed $\epsilon$ is stable when increasing the number of workers. Then the number of iteration in each node will decrease. 

For ECD-PSGD and mini-batch SGD, the effect is better when the gap is significant,  for synchronous first gain growth definition is always positive. When the gap is large, at the fixed iteration, the log loss from a particular algorithm worker setting is smaller.

In upper bound experiments, we use cost to indicate gain growth.

In our upper bound experiments, the iterations per worker can indicate the upper bound of scalability.

\subsection{Experiments Setting}
\subsubsection{Dataset}  
\begin{figure}[htbp]
	\includegraphics[width= \linewidth]{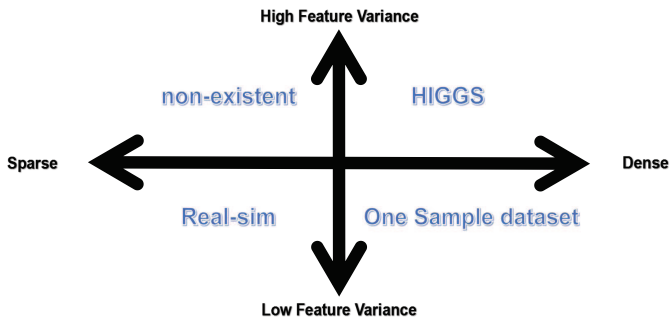}
	\caption{The best performance dataset for different algorithm}
	\label{axis_exp}
\end{figure}

We choose a sparse dataset with small feature variance as experiments dataset: real-sim dataset and a dense dataset with large feature variance: HIGGS dataset. The detail information about the above datasets is shown in table  \ref{dataset}. Their suitable algorithms are shown in figure \ref{axis_exp}.

In all cases, the dataset is randomly split into two parts: a training set containing 70\% of the dataset samples and a valid set containing 20\% of the dataset samples. 

\begin{table}[!htbp]
	
	\caption{Dataset information}
	\label{dataset}
	\begin{tabular}{|m{2cm}|c|c|c|c|}
		\hline
		dataset & \#features & \#size & feature range & density \\
		\hline
		real-sim & 20,958& 72,309&(0,1)&$<3$\%\\
		\hline
		HIGGS &28 &11,000,000 &(-4,3)&100\%\\
		\hline
		Simulated Data &20,958&-&0/1 & 70\% \\
		\hline
		Simulated Data: Small $LS_{\mathcal{A}}(\mathcal{D},\mathcal{S})$ dense dataset  &28/1000 &- &[-4,3] &100\%\\
		\hline
		Simulated Data: Large $LS_{\mathcal{A}}(\mathcal{D},\mathcal{S})$ dense dataset  &28/1000 &- &[-4,3] &100\%\\
		\hline
		Simulated Data: Small $LS_{\mathcal{A}}(\mathcal{D},\mathcal{S})$ sparse dataset  & 20,958 &- &[0,1] &$<3$\%\\
		\hline
		Simulated Data: Large $LS_{\mathcal{A}}(\mathcal{D},\mathcal{S})$ sparse dataset  & 20,958 &- &[0,1] &$<3$\%\\
		\hline
		Simulated Data: Low diversity dataset:  & 20,958 &72,309 &[0,1] &$<3$\%\\
		\hline 
		Simulated Data: middle diversity dataset:  & 20,958 &72,309 &[0,1] &$<3$\%\\        
		\hline 
		
	\end{tabular}
\end{table}

To match our theory, we also build three groups (nine in all) simulated datasets: (1)  Normal dataset for upper bound experiments    (2)  Different $LS_{\mathcal{A}}(\mathcal{D},\mathcal{S})$ datasets and (3) Different sample diversity datasets. The samples in those datasets are generated randomly and the label is generated by the function $label_i = sign(\xi_i\cdot ruler)$ where $ruler$ is the vector $(-1,2,-3,4,...,(-1)^{sample\_size}*sample\_size)$.

\textbf{ Small $LS_{\mathcal{A}}(\mathcal{D},\mathcal{S})$ dataset and large $LS_{\mathcal{A}}(\mathcal{D},\mathcal{S})$ dataset}
Small $LS_{\mathcal{A}}(\mathcal{D},\mathcal{S})$ dataset and large $LS_{\mathcal{A}}(\mathcal{D},\mathcal{S})$ dataset are used to match the $LS_{\mathcal{A}}(\mathcal{D},\mathcal{S})$ related theory.  All information is shown in table \ref{dataset}.

All samples' features in this group are sampled from the same distribution. In experiments, we use uniform distribution $\mathcal{U}(range\_begin, range\_end)$, where the range is shown in Table \ref{dataset}.

In $LS_{\mathcal{A}}(\mathcal{D},\mathcal{S})$ experiments, the size of the test dataset is 20\% of the number of training data. And the data in test data share the same feature distribution and density character with training data. 

In small $LS_{\mathcal{A}}(\mathcal{D},\mathcal{S})$   and dense dataset, the sample offered by $t$-th iteration is modified by the sample at $t-1$-th iteration: we randomly choose 10\% features and randomly change those features' value.  

In large$LS_{\mathcal{A}}(\mathcal{D},\mathcal{S})$  and dense dataset, the sample offered by $t$-th iteration is modified by the sample at $t-1$-th iteration: we randomly choose 90\% features and randomly change those features' value.

In small $LS_{\mathcal{A}}(\mathcal{D},\mathcal{S})$   and sparse dataset, the sample offered by $t$-th iteration is modified by the sample at $t-1$-th iteration: we randomly choose 10\% features and randomly change those features' value.  To make sample sparse, we also randomly pick some features and set them as zero, and the sparsity is equal to the sparsity of the sample at the first iteration.  

In large $LS_{\mathcal{A}}(\mathcal{D},\mathcal{S})$ and sparse dataset, the sample offered by $t$-th iteration is modified by the sample at $t-1$-th iteration: we randomly choose 90\% features and randomly change those features' value.  To make sample sparse, we also randomly pick some features and set them as zero, and the sparsity is equal to the sparsity of the sample at the first iteration.   

As we can see from the above dataset design setting, when two datasets' sparsity is the same, and the size of the dataset is large enough, two datasets are the same.

\textbf{Simulated Dataset for upper bound experiments}  Our experiment environment is poor, and it can only support 24 workers at once. 

For the upper bound of Hogwild!'s scalability on real-sim exceeds the number of cores of our computing environment. So we have to build a simulated dataset whose upper bound of scalability is easy to reach. In our simulated dataset, the density is 70\%. The feature distribution is the same as $LS_{\mathcal{A}(\mathcal{D},\mathcal{S})}$ experiments. Other information is shown in table \ref{dataset}.

Because of the limitation of the maximum number of workers, the DADM experiments also have to build a dataset to match the experiment results. In this experiment, we use randomly picked 1/8 real-sim dataset as our train dataset and use randomly picked 1/40 real-sim dataset as our test dataset.

In scalability upper bound experiments, the size of the test dataset is 20\% of the number of training data. Moreover, the data in test data only share the same feature range and density character with training data. 

\textbf{Simulated Dataset for sample diversity experiments} Measuring the sample diversity is a time costing jobs. The size of the dataset is not always positively correlated to the sample diversity (Considering one sample dataset).  Thus, we use the following method to build three different sample diversity datasets. The real-sim dataset is the high diversity dataset in the experiment.

We equally cut real-sim dataset into 4 part:

$\{  sub\_dataset_1, sub\_dataset_2, sub\_dataset_3, sub\_dataset_4 \}$

The sample diversity of sub\_dataset is lower than the whole dataset. Thus, we build the middle diversity dataset, abbr. real\_sim$_2$,  as follows:

$\{  sub\_dataset_1, sub\_dataset_1, sub\_dataset_2, sub\_dataset_2\}$

We also build the low sample diversity dataset, abbr. real\_sim$_4$ as follows:

$\{  sub\_dataset_1, sub\_dataset_1, sub\_dataset_1, sub\_dataset_1\}$

\subsubsection{Hardware}
We conducted our experiments on a server with 2 Xeon(R) CPU E5-2660 v2 @ 2.20 GHz, and 60G memory which contains twenty-four cores together.

%Because our experiment hardware is limited, we cannot conduct DADM experiments in our server: DADM requires that all samples load to memory at once, i.e., solve the subproblem minimum. Thus in the current version, we only present Hogwild!, mini-batch SGD, and ECD-PSGD's experimental results.

\subsubsection{Problem} In our experiment, we will solve the problem of training L2 norm logistic regression model because the log loss function suits all requirements which are asked by Hogwild!, mini-batch SGD, DADM, and ECD-PSGD. The logistic loss function is shown in Eq. \ref{experiment}.
\begin{align}
\mathop{argmin}_{x} \frac{1}{n}\sum_{i=1}^{n}\Phi(lable_i*\xi_i \cdot x) +\lambda/2\left\|x\right\|^2 
\label{experiment}
\end{align}
where $\Phi$ is the logistic loss, i.e., $\Phi(t) = log(1+e^{-t})$ and $\lambda = 0.01$. 

\subsection{Feature variance and Sparsity Experiment}
\subsubsection{Algorithm Setting}
In this experiment, we run HIGGS and real-sim on different algorithms to make the comparison.

In Hogwild!, the learning rate is $0.1$. In the mini-batch SGD and ECD-PSGD, learning rates are $0.1$. In Hogwild! experiments on the HIGGS dataset, to gain a stable curve, we have to set the mini-batch as 4. In the ECD-PSGD experiment, We connect all workers into a ring, and we do not compress the data. 

\subsubsection{Experimental Results}
\begin{figure}
	\centering
	\subfigure [HIGGS on mini-batch SGD]{   
		\includegraphics[width=0.7\columnwidth]{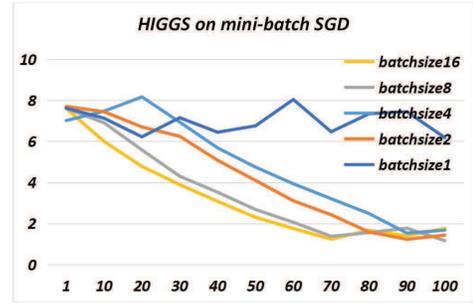} 
	} 
	\subfigure [Real-sim on mini-batch SGD] { 
		\includegraphics[width=0.7\columnwidth]{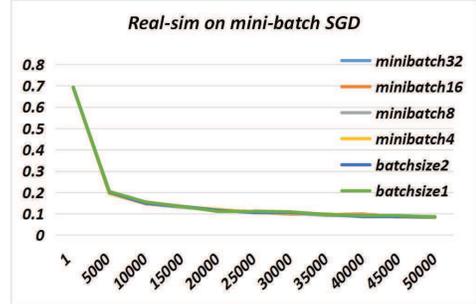} 
	} 
	\caption{ The performance of different datasets on mini-batch SGD. The X-axis is the number of iteration. Y-axis is test dataset log loss. In these cases,  the effect is better when the gap is larger: At the fixed iteration, the log loss from a particular algorithm worker setting is smaller. }
	\label{var_sparsity_mini} 
\end{figure}

\begin{figure}
	\centering
	\subfigure [Real-sim on ECP-PSGD]{ 
		\includegraphics[width=0.7\columnwidth]{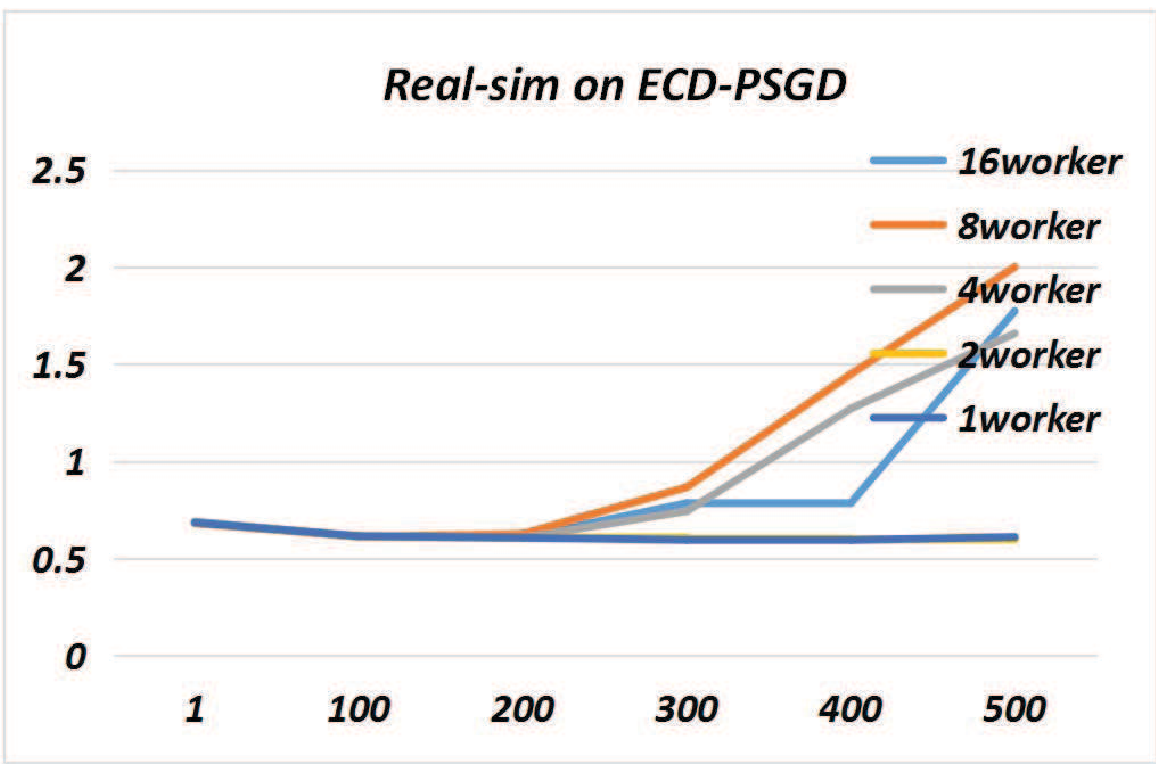} }
	\subfigure [HIGGS on ECD-PSGD]{ 
		\includegraphics[width=0.7\columnwidth]{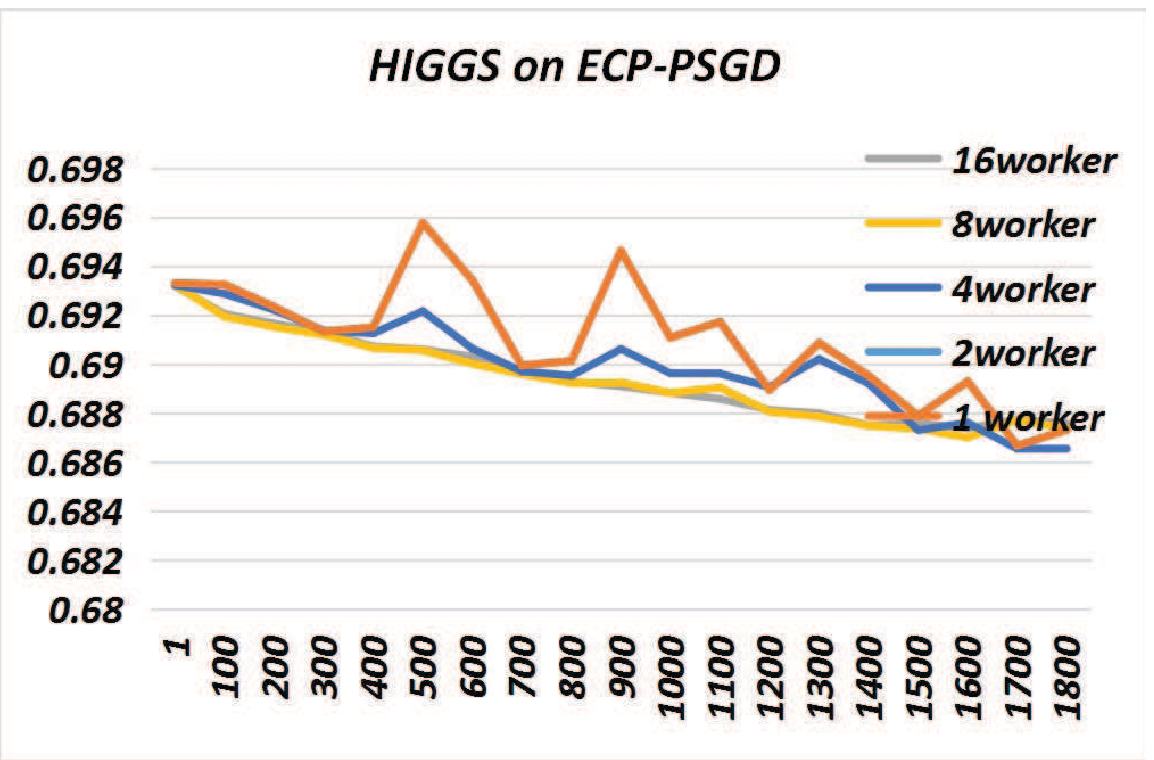} } 
	\caption{ The performance of different datasets on ECD-PSGD. The X-axis is the number of iteration. Y-axis is test dataset log loss. In these cases,  the effect is better when the gap is larger: At the fixed iteration, the log loss from a particular algorithm worker setting is smaller. } 
	\label{var_sparsity_ECD}
\end{figure}

\begin{figure}
	\centering
	\subfigure [Real-sim on Hogwild!]{ 
		\includegraphics[width=0.7\columnwidth]{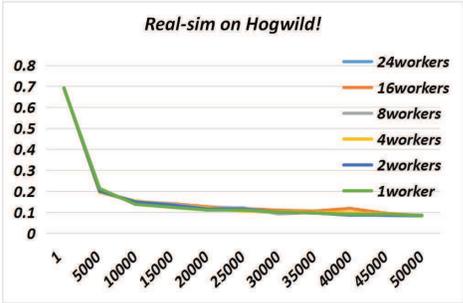} }
	\subfigure [HIGGS on Hogwild!]{ 
		\includegraphics[width=0.7\columnwidth]{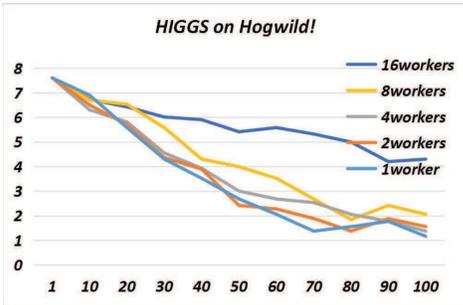} }  
	\caption{ The performance of different datasets on Hogwild!. The X-axis is the number of iteration. Y-axis is test dataset log loss. In these cases,  the effect is better when the gap is small: The number of iteration for the server to reach a fixed $\epsilon$ is stable when increasing the number of workers. Then the number of iteration in each node will decrease.  } 
	\label{var_sparsity_hog}
\end{figure}

The experimental results are shown in Figure \ref{var_sparsity_mini}, \ref{var_sparsity_hog}, \ref{var_sparsity_ECD}

\subsubsection{Experiment analysis}
In our feature variance and sparsity experiment, our experiment results well match to theory analysis: Our experiment results match the figure \ref{axis}. (1)In mini-batch SGD and ECD-PSGD, the parallel effect is markable for large variance dataset(HIGGS).  Large batch setting mini-batch SGD convergence faster, while for the sparse dataset(real-sim), the parallel technology does not exert any influence on convergence speed. For ECD-PSGD, parallel technology even brings a negative impact. (2) For the ASGD algorithm, i.e., Hogwild!, with the increasing number of workers, the influence on convergence speed is minor on the sparse dataset. The iteration number on each node is decreased linearly. However, for feature variance dataset(HIGGS), the convergence speed is drastically decreasing, which means the iteration number on each worker is not reduced obviously. In some cases, the iteration number on each worker is increasing with the number of worker's increasing.

\subsection{Sample Diversity Experiment}
\subsubsection{Algorithm Setting}

To present our experiments clearly, we use one worker to 16 workers to train real\_sim, real\_sim$_2$ and real\_sim$_4$ dataset. In each worker, the mini-batch size is one. 

\subsubsection{Experimental Results}
\begin{figure}
	\centering
	\subfigure  [Real-sim dataset on DADM] {  
		\includegraphics[width=0.7\columnwidth]{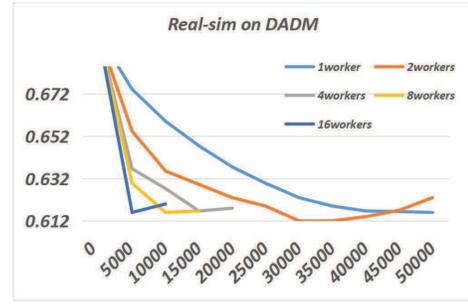} 
	} 
	\subfigure[Real-sim$_2$ dataset on mini-batch SGD] { 
		\includegraphics[width=0.7\columnwidth]{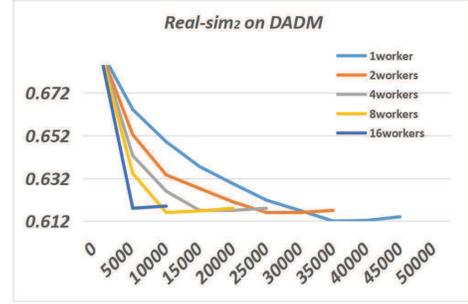} 
	}  
	\subfigure[Real-sim$_4$ dataset on mini-batch SGD] { 
		\includegraphics[width=0.7\columnwidth]{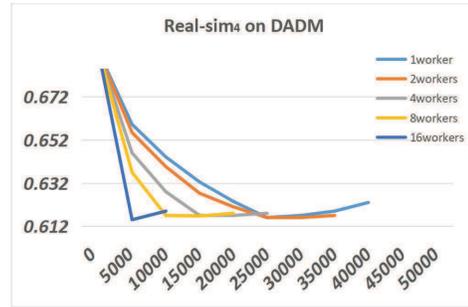} 
	} 
	\caption{ The performance of different sample diversity datasets on mini-batch SGD. The X-axis is the number of iteration. Y-axis is test dataset log loss. In these cases,  the effect is better when the gap is larger: At the fixed iteration, the log loss from a particular algorithm worker setting is smaller. }
	\label{DADM_div} 
\end{figure}

The experimental results are shown in Figure \ref{DADM_div}.

\subsubsection{Experiment analysis}
In our sample diversity experiment, our experiment results well match to theory analysis: high sample diversity leads to better scalability. In DADM, when sample diversity is large, at the same iteration, the more gain growth we can get: the gap between the different lines is large.

\subsection{$LS_{\mathcal{A}}(\mathcal{D},\mathcal{S})$ Experiment}
\subsubsection{Algorithm Setting}
The algorithm setting in this section is the same as the feature variance and sparsity section. The above sections show that different datasets suit different algorithms; we only present 1. Sparse dataset for Hogwild! 2. Feature variance dataset for mini-batch SGD (\#feature is 28) and ECD-PSGD (\#feature is 1000). In Hogwild! and DADM experiment, the first sample is sampled from the real-sim dataset. In the mini-batch SGD experiment, the first sample is sampled from the HIGGS dataset. In the ECD-PSGD experiment, we use our first sample to make the gap between different curves large, and the size of this sample is 1000.

\subsubsection{Experimental Results}
\begin{figure}
	\centering
	\subfigure  [Large $LS_{\mathcal{A}}(\mathcal{D},\mathcal{S})$ dataset on mini-batch SGD] {  
		\includegraphics[width=0.7\columnwidth]{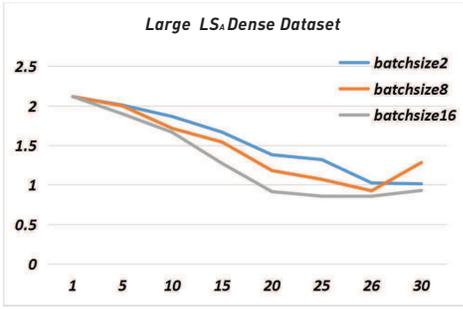} 
	} 
	\subfigure[Small $LS_{\mathcal{A}}(\mathcal{D},\mathcal{S})$ dataset on mini-batch SGD] { 
		\includegraphics[width=0.7\columnwidth]{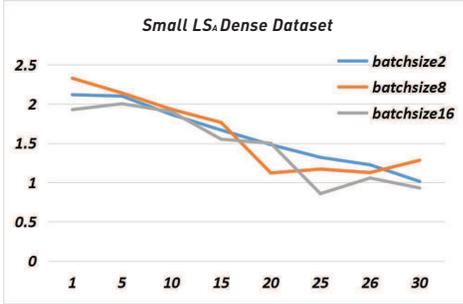} 
	}  
	\caption{ The performance of different $LS_{\mathcal{A}}(\mathcal{D},\mathcal{S})$ dataset on mini-batch SGD. X-axis is the number of iteration. Y-axis is test dataset logloss. In this cases,  the effect is better when the gap is larger: At the fixed iteration, the log loss from a particular algorithm worker setting is smaller. }
	\label{c_sim_exp_mini} 
\end{figure}

\begin{figure}
	\centering
	\subfigure[Large $LS_{\mathcal{A}}(\mathcal{D},\mathcal{S})$ dataset on ECD-PSGD]{  
		\includegraphics[width=0.7\columnwidth]{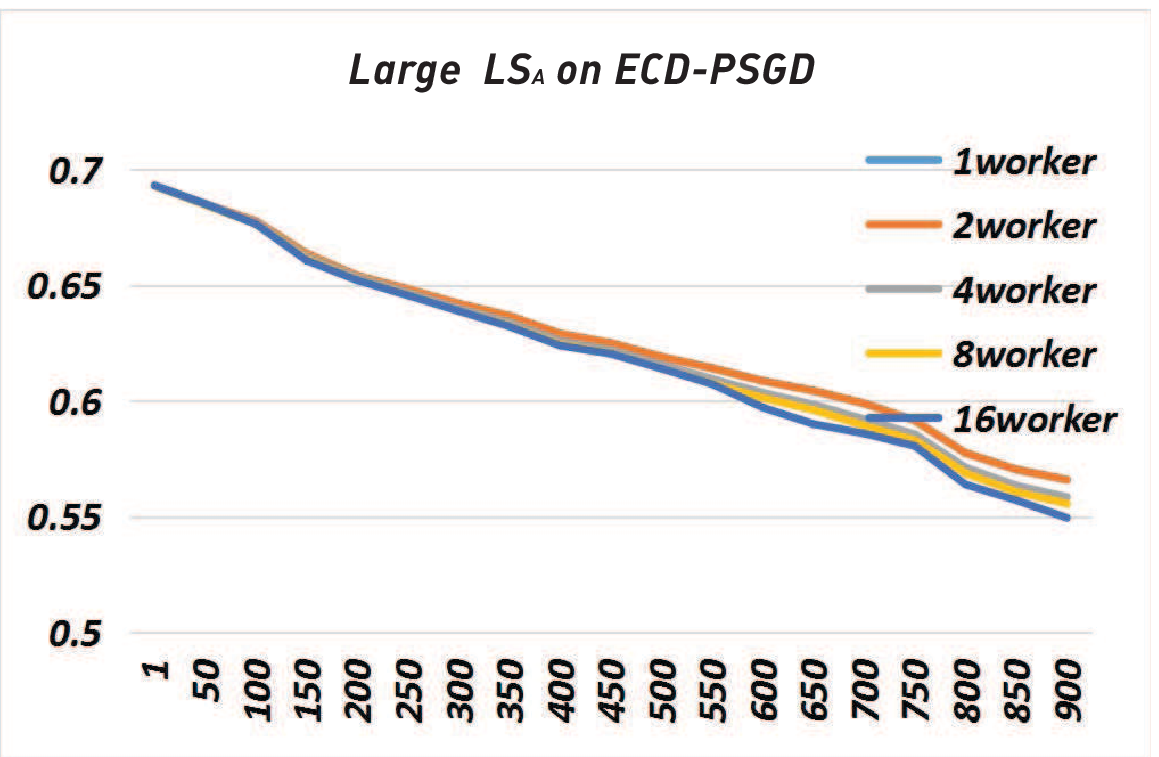} 
	} 
	\subfigure [small $LS_{\mathcal{A}}(\mathcal{D},\mathcal{S})$ dataset on ECD-PSGD]  { 
		\includegraphics[width=0.7\columnwidth]{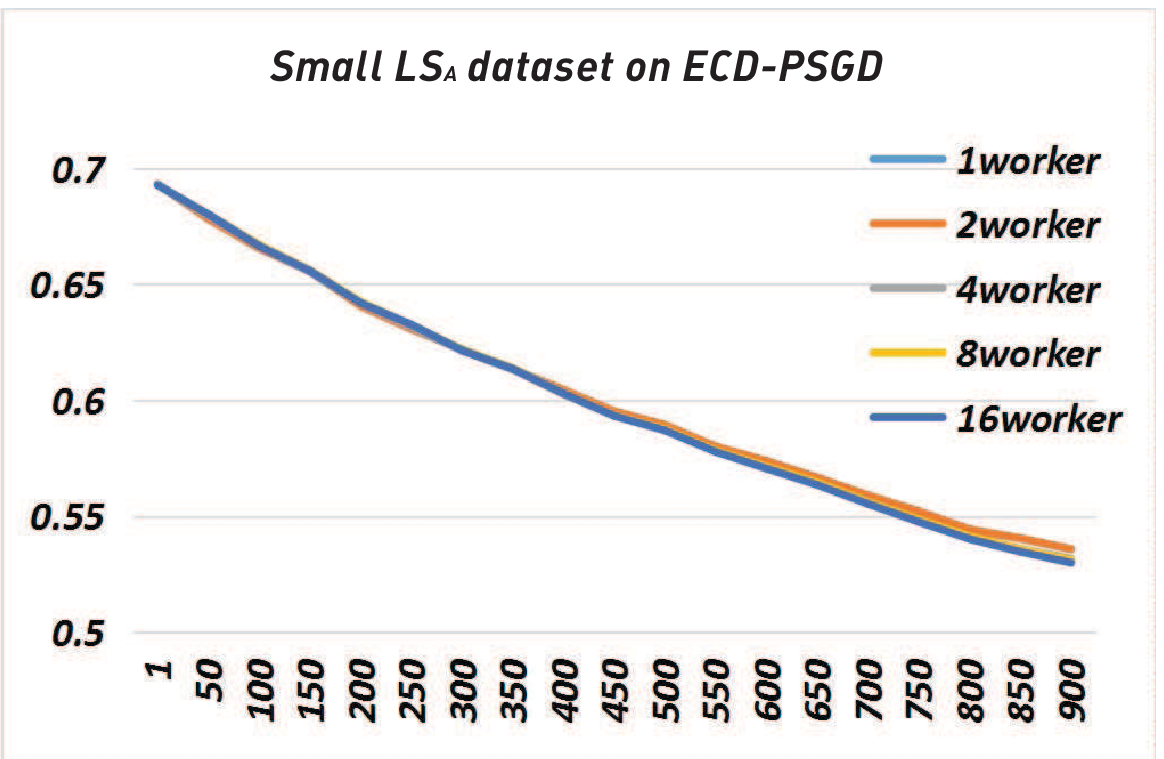} 
	}  
	\caption{ The performance of different $LS_{\mathcal{A}}(\mathcal{D},\mathcal{S})$ dataset on ECD-PSGD. X-axis is the number of iteration. Y-axis is test dataset logloss. In this cases,  the effect is better when the gap is larger: At the fixed iteration, the log loss from a particular algorithm worker setting is smaller. } 
	\label{c_sim_exp_ECD}
\end{figure}

\begin{figure}
	\centering
	\subfigure [Large $LS_{\mathcal{A}}(\mathcal{D},\mathcal{S})$ dataset on Hogwild!]{ 
		\includegraphics[width=0.7\columnwidth]{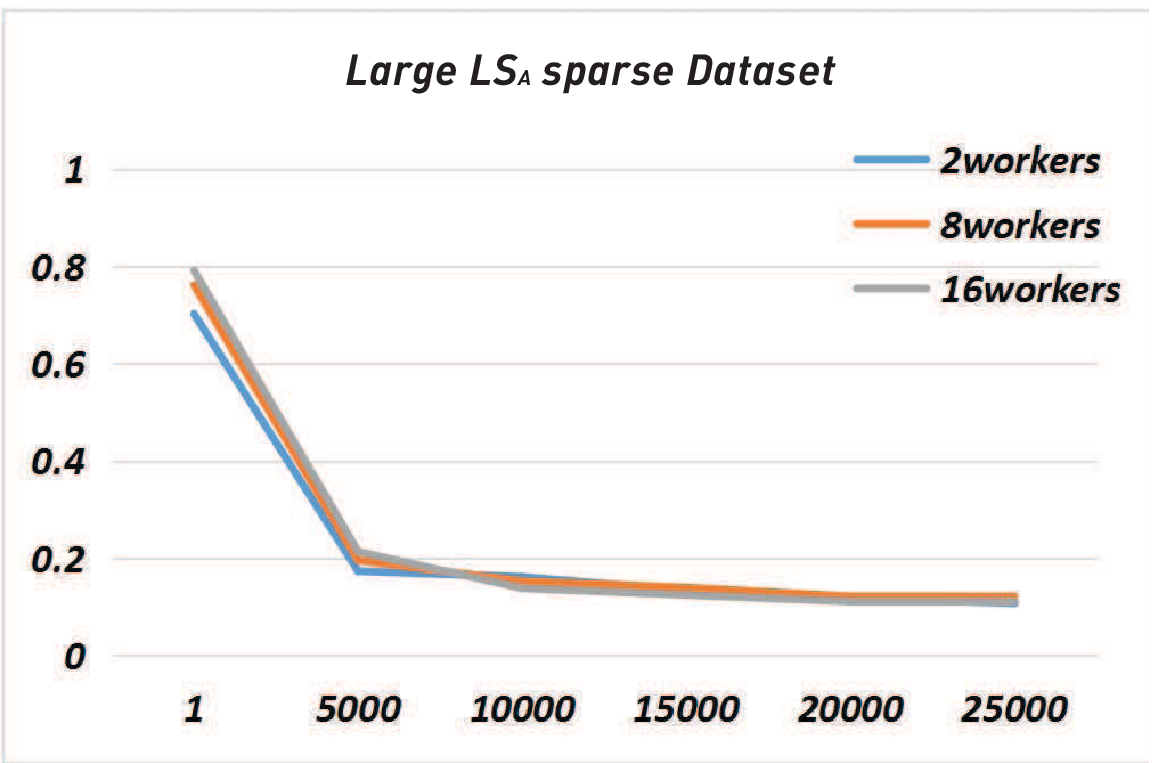} 
	} 
	\subfigure [Small $LS_{\mathcal{A}}(\mathcal{D},\mathcal{S})$ dataset on Hogwild!]{  
		\includegraphics[width=0.7\columnwidth]{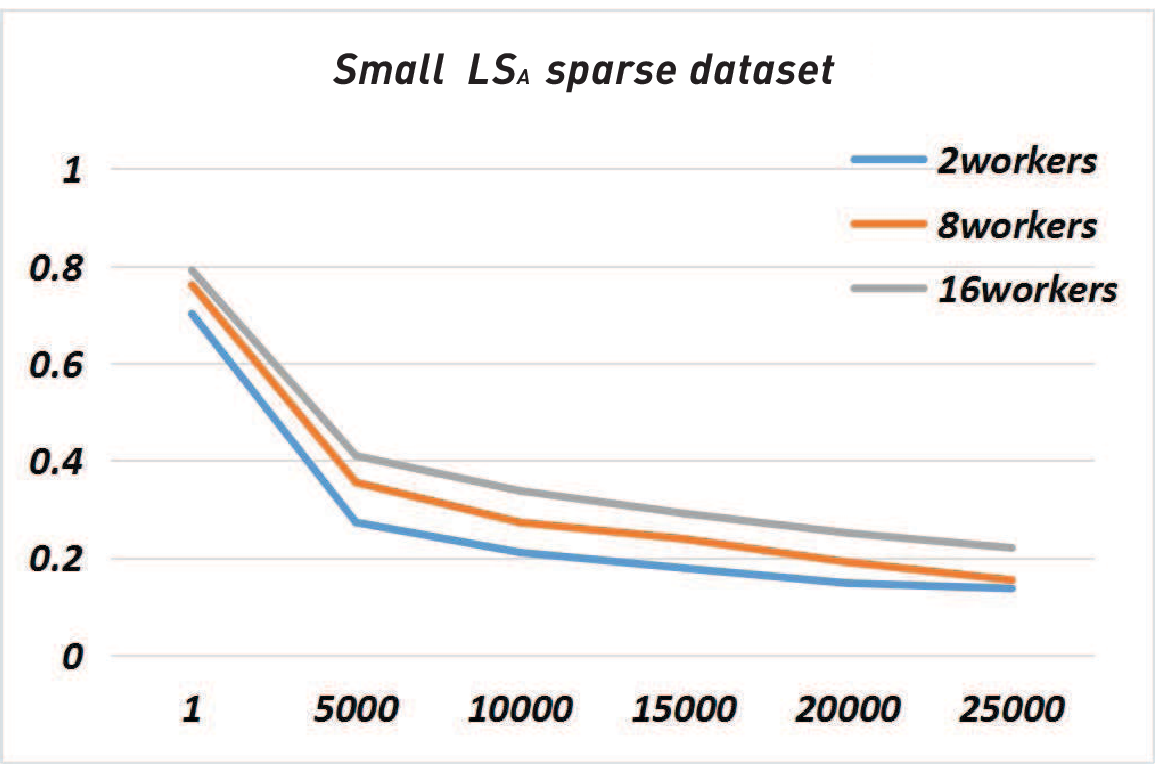} 
	}  
	\caption{ The performance of different  $LS_{\mathcal{A}}(\mathcal{D},\mathcal{S})$ dataset on Hogwild!. The X-axis is the number of iteration. Y-axis is test dataset log loss. In these cases,  the effect is better when the gap is small: The number of iteration for the server to reach a fixed $\epsilon$ is stable when increasing the number of workers. Then the number of iteration in each node will decrease.  } 
	\label{c_sim_exp_hog}
\end{figure}

The experimental results are shown in Figure \ref{c_sim_exp_mini}, \ref{c_sim_exp_hog},\ref{c_sim_exp_ECD}.

\begin{figure}
	\centering
	\subfigure [Large $LS_{\mathcal{A}}(\mathcal{D},\mathcal{S})$ dataset on DADM]{ 
		\includegraphics[width=0.7\columnwidth]{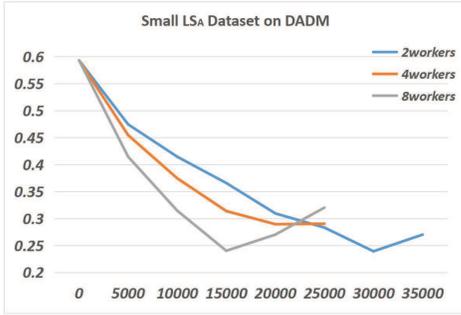} 
	} 
	\subfigure [Small $LS_{\mathcal{A}}(\mathcal{D},\mathcal{S})$ dataset on DADM]{  
		\includegraphics[width=0.7\columnwidth]{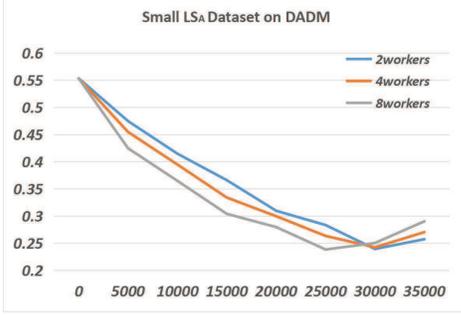} 
	}  
	\caption{ The performance of different  $LS_{\mathcal{A}}(\mathcal{D},\mathcal{S})$ dataset on DADM. X-axis is the number of iteration. Y-axis is test dataset logloss. In this cases,  the effect is better when the gap is large: At the fixed iteration, the log loss from a particular algorithm worker setting is smaller.  } 
	\label{c_sim_exp_dadm}
\end{figure}

\subsubsection{Experiment analysis}
In our $LS_{\mathcal{A}}(\mathcal{D},\mathcal{S})$ experiment, our experiment results well match to theory analysis: large $LS_{\mathcal{A}}(\mathcal{D},\mathcal{S})$ value leads to better scalability. In mini-batch SGD, DADM and ECD-PSGD, when  $LS_{\mathcal{A}}(\mathcal{D},\mathcal{S})$ is large, at the same iteration, the more gain growth we can get: the gap between the different line is large. For ASGD algorithm, i.e. Hogwild!,  when $LS_{\mathcal{A}}(\mathcal{D},\mathcal{S})$ is large, the more gain growth we can get: the gap between the different line is small, which means each worker trains fewer iterations.

\subsection{Scalability Upper Bound Experiment}
\subsubsection{Algorithm Setting}

The algorithm setting in this section is the same as the feature variance and sparsity section. The above sections show that different datasets suit different algorithms, we only present: 1. sparse dataset for Hogwild! 2. feature variance dataset for mini-batch SGD and ECD-PSGD. 

Our experiment environment cannot reach the upper bound of scalability of the real-sim dataset: Our experiment environment only supports twenty-four thread (worker) in all. Thus, in Hogwild! experiment, we have to use a simulated dataset. In the mini-batch SGD and ECD-PSGD experiment, we use the HIGGS dataset. 

\subsubsection{Experimental Results} 
The results is shown in table \ref{indx}. The upper bound is between two red marked values. 
\begin{table}[!htbp]
	\centering
	\caption{The  iteration per worker for different algorithm.}
	\label{indx}
	\begin{tabular}{|m{1cm}|c|c|c|c|c|}
		\hline
		Algorithm  & 2workers & 4workers & 8workers & 16workers &24workers \\
		\hline
		Hogwild!  & 376 & 321 & \textcolor{red}{\textbf{356}} & \textcolor{red}{\textbf{412}} & - \\
		\hline
		mini-batc &91 & 87 & \textcolor{red}{\textbf{71}} &\textcolor{red}{\textbf{69}} &-\\
		\hline
		ECD-PSGD  &\textcolor{red}{\textbf{1654}} & \textcolor{red}{\textbf{1621}} & 1623&1648 &-\\
		\hline
		DADM &22596 &11421 & 6358 & \textcolor{red}{4064} & \textcolor{red}{3972}\\
		%DADM &{22596} &11421 &6258‬& \textcolor{red}{\textbf{4064}}& \textcolor{red}{\textbf{3972}}\\
		\hline    
	\end{tabular}
\end{table}

\subsubsection{Experiment analysis}
In table \ref{indx}, we show that the different algorithms have their upper bound scalability, which is marked by red in table \ref{indx}, even using their best performance dataset. Based on our analysis in "The Upper Bound of Scalability" Section, the growth gain for Hogwild! is negative. For ECD-PSGD and mini-batch SGD, the growth is close to zero. Thus, in the range which we marked, the algorithms meet their scalability upper bounds.

\subsection{Some notes in experiment}
For some readers always put some questions on experiments, we put some notes in this subsection. Our experiments are well-designed instead of grasping or building some datasets.  

\subsubsection{Why not time measure}
In brief, using time as the x-axis is inessential, and the data of time can be easily overstated.

Using time as an x-axis is inessential because of the PCA.

Because of the PCA and the mapping relationship between the number of iteration and time in the perfect computer,  iteration can be mapped into time in a perfect computer cluster directly.  What is more,  the upper bound of scalability performance of an algorithm on a real cluster is the scalability of the performance of this algorithm on the perfect computer cluster.

The data of time can be easily overstated. 

Time is a comprehensive index that can be influenced by many factors like the hardware and the size of the machine learning model. The following example shows that the size of the model and the network hardware exert great influence on scalability. We can gain any scalability result in a deliberately designed compute environment.

\begin{example}
	When using time as x-axis and training LR machine learning model, the scalability of Hogwild! on KDD 2012 dataset on the cluster, which is connected by Intel Corporation I350 Gigabit Network Connection, is worse than ECD-PSGD with appropriate compression.     However, when using time as x-axis and training LR machine learning model, the scalability of Hogwild! on KDD 2012 dataset on the Era supercomputer is better than ECD-PSGD with whatever compression.
\end{example}

In the above example, the model size is large (427MB).  Intel Corporation I350 Gigabit Network Connection network card needs more time to convey the whole model in the network. The bottleneck is the network. Thus, ECD-PSGD, using appropriate compression to reduce the burden of the network, performs better than Hogwild!. However, when the algorithms run on the platform with Infiniband, the impact of network hardware is reduced. Thus, Hogwild! performs better than ECD-PSGD.

Thus, in experiments, we only show the convergence curve on the figure whose X-axis is the number of iteration, and Y-axis is the log loss.

\subsubsection{Why $x^*$ in our experiments is worse than others}
We do not want to compare which algorithm can find the $x^*$ better. We want to compare the scalability of the algorithm. Thus, we do not fine-grained control the algorithm parameters, like the learning rate.

\subsubsection{Why some experiments using simulated dataset}
(1) Some statistical value is hard to control and our limitation of computing resource. 

In local similarity experiments, it is costly to build two sequences that have remarkably different $LS_{\mathcal{A}}$ in one dataset.  In diversity experiments,  calculating diversity for each dataset is also a time costing job. 

(2)  It is appropriate to tease out the problem. It is hard to find two datasets which are almost the same without target characters, which are used in control variant experiments. Simulated datasets can build almost the same datasets, and these datasets are only different in target characters.  Control variant experiments would help experiments match theories.

Thus, we have to build our own experiments datasets.

\section{Conclusion and Discuss}

\subsection{Conclusion}
Based on our analysis and experiments, we can draw the following conclusion clearly: 1. Different datasets suit different parallel stochastic optimization algorithm. 2. Before training a machine learning model, rearrange dataset is an ideal choice. 3. No matter which parallel stochastic optimization algorithm is, there always exists an upper bound of scalability. 4. The scalability performance for certain datasets on a specific machine learning model cannot be pushed into other cases.

It is worthy to note that some machine learning models, like CNN or DNN,  do not obey the convex, Lipschitz, or continuity requirements. Thus, the scalability of the algorithm on those models need to be analyzed.

\section{Acknowledgment}
This work was supported by the National Natural Science
Foundation of China under Grant No. 61432018, Grant No.
61502450, Grant No. 61521092, and Grant No. 61272136, and by the
National Major Research High-Performance Computing
Program of China under Grant No. 2016YFB0200800. State Key Laboratory of Computer Architecture Foundation under Grant No. CARCH3504

This work is finished during the Cheng Daning and Zhang Hanping's internship in Wisdom Uranium technology Co.Ltd.

Corresponding author is Shigang Li
\bibliographystyle{IEEEtran}
\bibliography{IEEEabrv,mybibfile}

\newpage

\section*{Appendix}
\subsection{Algorithm Description}
\subsubsection{Notes and Symbols}
To make our present clearly, we summarize the algorithm descriptions common notes and symbols here. $n$ is the number of sample in dataset. $m$ is the number of worker. $\gamma$ is the learning rate. $\lambda$ is the regularization coefficient. $G_{\xi_i}(x)$ and $\nabla F(x;\xi_i)$ are the sub-gradient of function $F(x;\xi_i)$. To make reader easy to match the algorithm descriptions in their original paper, we keep them all in our algorithm descriptions.  $Q$ is the collection of samples which are in a mini-batch.  $batch\_size$ is the number of $Q$ and $local\_batch\_size$ is the number of $Q_{local}$ which is the mini-batch in a worker.

\subsubsection{Hogwild!}
Hogwild! is the most important asynchronous parallel SGD algorithm. Hogwild! is the base of the current machine learning frame: Parameter Server framework.

The Algorithm \ref{Hogwild!} is the description of Hogwild!. It is worthy of mentioning that $F(x;\xi)$ is not the loss function directly. $F(x;\xi)$ should be written as hypergraph form\cite{Niu2011HOGWILD}. 
\begin{algorithm}[thbp]
	\begin{algorithmic}
		\STATE \textbf{In:} $1$ Server, $m$ worker, random delay $\tau$ ( $0<\tau<\tau_{max}$), learning rate $\gamma$
		\STATE \textbf{Out:} $x^*$, which is the argmin of $f(x) $
		\STATE 
		\STATE \textbf{WORKER:}
		\REPEAT
		\STATE 1. Pick sample $\xi_i$ from dataset;
		\STATE 2. Pull Model $x_i$ from Server;
		\STATE 3. Compute $G_{\xi_i}(x_i)$, which is the sub-gradient of $F(x;\xi_i)$
		\STATE 4. Push $G_{\xi_i}(x)$ into Server.
		\UNTIL{Forever}
		\STATE 
		\STATE \textbf{SERVER:}
		\REPEAT
		\STATE 1. Receive $G_{\xi_i}(x_{j-\tau})$ from any worker. 
		\STATE 2. $x_{j+1}$ = $x_j$ + $\gamma$$G_{\xi_i}(x_{j-\tau
		})$
		\UNTIL{Forever}
		\STATE    3. Return $x^*$
	\end{algorithmic}
	\caption{Hogwild!}
	\label{Hogwild!}    
\end{algorithm}

\subsubsection{Mini-batch SGD algorithm}

Mini-batch SGD algorithm is the most critical data-parallel SGD algorithm. Nowadays, mini-batch SGD is the main parallel method that is implemented in the supercomputer.

Algorithm \ref{Batch SGD} is the description of the mini-batch SGD algorithm.

\begin{algorithm}[htbp]
	\begin{algorithmic}
		\STATE \textbf{In:} $1$ Server, $batch\_size$ Workers, learning rate $\gamma$
		\STATE \textbf{Out:} $x^*$, which is the argmin of $f(x)$
		\STATE
		\STATE \textbf{WORKER:}
		\FOR{Forever}{
			\STATE 1. Pick sample $\xi_i$ from dataset;
			\STATE 2. Receive $x_i$ from Server
			\STATE 3. Compute $G_{\xi_i}(x_i)$, which is the sub-gradient of $F(x_i;\xi_i)$
			\STATE 4.Push $G_{\xi_i}(x)$ into Server.
		}
		\ENDFOR
		\STATE 
		\STATE \textbf{SERVER:}
		\FOR{Forever}{
			\STATE 1. All-gather $G_{\xi_1}(x_j),G_{\xi_2}(x_j),...,G_{\xi_{batch\_size}}(x_j)$ from $worker_1,worker_2,,,,,worker_{batch\_size}$
			\STATE 2. Compute  $G_{ave}(x_j) = \frac{1}{batch\_size} \sum_{i=1}^{batch\_size}G_{\xi_i}(x_j)$
			\STATE 3. $x_{j+1}$ = $x_j$ + $\gamma$$G_{ave}(x_j)$
		}\ENDFOR
		\STATE    4. Return $x^*$
		
	\end{algorithmic}
	\caption{Mini-batch SGD algorithm}
	\label{Batch SGD}    
\end{algorithm}

\subsubsection{Distributed Alternating Dual Maximization(DADM)}
DADM\cite{Zheng2017A} depends on the dual ascent method to gain a minimum of $f(x)$. The DADM can be treated as the mini-batched SDCA algorithm. DADM selected an intermediate variable to help different components of mini-batch are computed in the different nodes in a cluster.

The full version of DADM can be complex, and it tries to solve the goal function, which contains three parts. However, when it comes to the common machine learning problem, the algorithm is presented in a simple form, like algorithm \ref{DADM}. In algorithm \ref{DADM}, $L(x;\xi)$ is the loss function. $L^*$ and $\psi^*$ is the convex conjugate function of $F$ and $\psi$. $\alpha_i$ is the dual variables. To make our present clearly, we omit some explanations. Some notes are different from the original algorithm description\cite{Zheng2017A}. Again, in this paper, our target is not showing every detail of the algorithm. We focus on algorithm scalability performance.

\begin{algorithm}[htbp]
	\begin{algorithmic}
		\STATE \textbf{In:} $1$ Server, $m$ Workers, $batch\_size = n * local\_batch\_size$ , learning rate $\gamma$, $\alpha_i$=$v^0$=0 
		\STATE \textbf{Out:} $x^*$, which is the argmin of $f(x)$
		\STATE
		\STATE \textbf{WORKER:}
		\FOR{t=1,2,...,forever}{
			\STATE 1. Pick $local\_batch\_size$ samples as $Q_{local}$, $\xi_{j_1}$, $\xi_{j_2}$ ... $\xi_{j_{local\_batch\_size}}$$\in Q_{local}$ from dataset;
			\STATE 2. Receive $\Delta v^{t-1}$ from Server
			\STATE 3. $v^t_{local} = v^{t-1}_{local}+\Delta v$
			\STATE 4. Approximately maximize Eq. \ref{DADM_eq},w.r.t $\Delta \alpha_i$
			\begin{align}
			\Delta \alpha_{Q_{local}} &= \mathop{argmin}_{\alpha_{Q_{local}}}  \sum_{i\in Q_{local}} -L^*(-\alpha^{t-1}_i-\Delta \alpha_i) \notag\\
			&-\lambda \psi^*(v^{t-1}_{local}+\frac{\sum_{i \in Q_{local}}\xi_i\cdot\Delta \alpha_i}{\lambda n/m})
			\label{DADM_eq}
			\end{align}
			\STATE 5. Send $\Delta v_{local}^{t} = \frac{1}{\lambda}\sum_{i \in Q_{local}}\xi_i \cdot \Delta \alpha_i$
		}
		\ENDFOR
		\STATE 
		\STATE \textbf{SERVER:}
		\FOR{t=1,2,...,Forever}{
			\STATE 1. All-gather $\Delta v_{local\_i}^t$ from $worker_i(i=1,2...,m)$ 
			\STATE 2. Compute  $\Delta v^t = \frac{1}{n}\sum_{i=1}^{m}\Delta v_{local\_i}^t$
			\STATE 3. Broadcast $\Delta v^t$ to all workers
		}\ENDFOR
		\STATE    4. Return $x^* = \nabla \psi^*(v)$ 
		
	\end{algorithmic}
	\caption{DADM}
	\label{DADM}    
\end{algorithm}

\subsubsection{ECD-PSGD} 
Decentralization and compression stochastic gradient methods are a new hot topic. To reduce the burden of the network, different workers send compressed information to neighborhood workers. Then, they average their models.

We choose one of the states of the art decentralization and quantization SGD algorithm: ECD-PSGD \cite{tang2018decentralization} as our example. In ECP-PSGD, we will show how datasets influence the algorithm scalability.

The description of ECP-PSGD is shown in the algorithm \ref{ECD-PSGD}. Again, we still omit some explanations. We only offer a basic version of the ECD-PSGD algorithm: all nodes share the same amount of data, and all nodes share the same weight. In this algorithm description, $x^{(i)}$ is the model in $ith$ worker. The worker weight and network are described by matrix $W$. $W_{i,j}$ is the element in $W$'s $i$ row and $j$ column and  $1=\sum_{i=1}^{m}W_{i,j} = 1$. The connected neighbors of one worker $i$ here refer to all workers that satisfy $W_{i,j}\neq 0$. 

\begin{algorithm}[htbp]
	\begin{algorithmic}
		\STATE \textbf{In:} $m$ Workers, Weighted and network matrix $W$, learning rate $\gamma$, initial point $x^{i}_1 = x_0$, initial intermediate variable $y^{(i)} = x_0$
		\STATE \textbf{Out:} $x^*$, which is the argmin of $f(x)$
		\STATE
		\STATE \textbf{WORKER:}
		\FOR{t=1,2,...,forever}{
			\STATE 1. Pick a sample $\xi_t$ from dataset;
			\STATE 2. Compute a local stochastic gradient based on $\xi_i$: $\nabla F(x^{(i)}_t;\xi_t)$
			\STATE 3. Pull compressed $y^{(j)}$ as  $\hat{y}^{(j)}$ from neighbors worker and compute
			\begin{equation*}
			x_{t+\frac{1}{2}}=\sum_{j=1}^{m}W_{i,j}\hat{y}_t^{(j)}
			\end{equation*}
			\STATE Update local model
			\begin{equation*}
			x_{t+1}=x_{t+\frac{1}{2}} - \gamma  \Delta F(x^{(i)}_t;\xi_t)
			\end{equation*}
			\STATE 4. Each worker compute the z-value of itself:
			\begin{equation*}
			z^{(i)}_{t+1}=(1-t/2)x^{(i)}_t+\frac{t}{2}x^{(i)}_{t+1}
			\end{equation*} 
			and compress $z^{(i)}_{t+1}$ into $C(z^{(i)}_{t+1})$
			\STATE 5. Each worker update intermediate variable for its connected neighbors:
			\begin{equation*}
			y^{(i)}_{t+1}=(1-2/y)y^{(i)}_t+\frac{2}{t}C(z^{(i)}_{t+1})
			\end{equation*}
		}
		\ENDFOR
		\STATE 6. Output:$x^*=\frac{1}{m}\sum_{i=1}^{m}x^{(i)}$
	\end{algorithmic}
	\caption{ECD-PSGD}
	\label{ECD-PSGD}    
\end{algorithm}

\subsection{Theorem Analysis}
To make our presentation clearly, we omit non-relevant parameters for those following lemmas and theorems in later parts. In the following part, $h_i(\cdot)$ are the functions which only contains the parameters which are related to the machine learning model, initial value $x_0$ and algorithm parameter like $\lambda$ and $\gamma$. $h_i(\cdot)$ do not care about the character of datasets and how many nodes we will use, i.e., the value of $m$.
\subsubsection{Hogwild!}
Firstly, we present a necessary conclusion which builds the connection between the number of workers and the lag(delay) between when a gradient is computed and when it is used in the Parameter Server Framework.

\begin{theorem}
	The minimum of the maximum of $\tau$ is the number of workers, i.e., $m \leq \tau_{max}$.  And when all workers share the same performance, the system would achieve the minimum.
\end{theorem}

The convergence analysis of Hogwild! is shown in theorem \ref{hog_theorem}. This theorem is transformed theorem from the Niu et al. 's work\cite{Niu2011HOGWILD}.
\begin{theorem}
	Suppose in algorithm \ref{Hogwild!} that the lag, i.e. $\tau$, which is between when a gradient is computed and when it is used, is always less than or equal to $\tau_{max}$, and $\gamma$ is under certain condition. for some $\epsilon > 0$.When $t$ is an integer satisfying
	\begin{equation*}
	t\geq (1+6\tau_{max}\rho+6\tau_{max}^2\Omega\delta^{1/2})\Omega h(\epsilon)
	\end{equation*}
	Then after $t$ component updates of $x$, we have $\mathbb E [f(x_t)-f(x^*)]<\epsilon$.
	$h(\epsilon)$ is only influenced by the character of $f(\cdot)$ and initial value $x_0$.
	\label{hog_theorem}
\end{theorem}

In theorem \ref{hog_theorem}, $\rho$ is the probability that any two $G_{\xi_i}(x_i)$ and $G_{\xi_j}(x_j)$ have the same nonzero value at the same feature; $\Omega$ is the max number of nonzero feature in $G_{\xi}(x)$; $\delta$ is simply the maximum frequency that any feature appears in $G_{\xi}(x)$.

\textbf{Sparsity and Feature variance} As we can see, when each worker shares the same performance, each worker needs to train $t/m = (1/m+6\rho+6m\Omega\delta^{1/2})\Omega h(\epsilon)$ which means with the increasing of the number of workers, each worker may have to exert more iterations. To make each workers training less iteration with increasing the number of workers, the $\Omega\delta^{1/2}$ should be extremely small: When $m$ is large enough, we expect that $ 1/(m+1)+6(m+1)\Omega\delta^{1/2} <  1/m+6m\Omega\delta^{1/2}$, which means we can gain benefit when we use more resource, i.e., a good algorithm scalability. The above facts show that the scalability of Hogwild is controlled by the value of $\Omega\delta^{1/2}$. 

When we decide which machine learning model we use, the sparsity of the dataset is the only factor that influences the $\Omega$ and $\delta$. From the definition of $\Omega$, $\delta$ and $\rho$, we can gain conclude that $\Omega,\delta$ and the sparsity of $G_{\xi}(x)$ is a positive correlation. For the common machine learning model, like SVM, LR, neural network, the relationship between the sparsity of samples in a dataset and the sparsity of $G_{\xi}(x)$ is clearly and significantly positive correlation. Especially, when machine learning models are linear models like SVM and LR, the sparsity of $G_{\xi_i}(x)$ is equal to the sparsity of $\xi_i$. 

The above conclusion is also shown in other ASGD algorithms convergence analysis like delay-tolerate ASGD and quantization ASGD.

Theorem \ref{hog_theorem} shows that feature variance plays no influence on algorithm scalability. However, when the dataset is sparse, the feature variance must be low: for any feature, in most samples in the dataset,  this feature is zero. 

\textbf{The influence of $LS_{\mathcal{A}}(\mathcal{D},\mathcal{S})$}
The influence of $LS_{\mathcal{A}}(\mathcal{D},\mathcal{S})$ is buried in the proof of theorem \ref{hog_theorem}. The conclusion is that $LS_{\mathcal{A}}(\mathcal{D},\mathcal{S})$ is positively correlated to the scalability. The proof of this part we put in Appendix part for this part needs to cite a lot of proof context from the work\cite{Niu2011HOGWILD}.

\textbf{The upper bound of scalability} From theorem \ref{hog_theorem}, we draw the scalability upper bound, which is decided by the character of the dataset. To make time faster, at least each worker should train less sample compared with one worker, i.e., $ 1/m+6m\Omega\delta^{1/2} <  1/1+6*1*\Omega\delta^{1/2}$. However, the function $constant_1x+constant_2/x$ ($constant_1, constant_2>0$) is increasing function when $x$ is large enough. Thus the maximum of $m$, which satisfies $ 1/m+6m\Omega\delta^{1/2} <  1/1+6*1*\Omega\delta^{1/2}$ is the maximum number of worker we can use in Hogwild!. The upper bound of Hogwild! scalability suits the second situation in "The Upper Bound of Algorithm" section.

\subsubsection{Mini-batch SGD}
Again, we present the basic fact which builds the connection with the degree of parallelism and batch size. The following fact is valid.

\begin{fact}
	In algorithm \ref{Batch SGD}, the upper bound of the number of workers is the batch size.
\end{fact}

To make our presentation clear, we show our theorem about the convergence of the mini-batch SGD algorithm:
\begin{theorem}
	When goal function Eq. \ref{reg_goal} is running on algorithm \ref{Batch SGD}, then we have
	\begin{align*}
	&\mathbb{E}_{x_t\in D_t}f(x_t) - f(x^*) \leq \notag\\
	&\bigg((h_2\left(F\left(\cdot\right)\right)\left(d(\mu_{D^t},\mu_{D^*})+\frac{\sigma_{D^*}+W_2(D^0,D^*)(1-\gamma\lambda)^t}{(batch\_size)^{t/2}}\right)\\
	&+h_3\left(F\left(\cdot\right)\right)\bigg)^2
	\end{align*}
	where $D^t$ is the distribution of $x^t$, $D^*$ is the distribution of $x^*$, $\sigma_{D}$ is the standard deviation of distribution $D$, $\mu_D$ is the mean of distribution $D$. $W_2(D_1,D_2)$ is the Wasserstein metrics between $D_1$ and $D_2$.
	\label{batch_sgd_the}
\end{theorem}

\textbf{Sparsity and Feature variance} When dataset and machine learning model are chosen, $D^0$ and $D^*$ would be determined. For most of the cases, the $x^*$ is a fixed number. The value of $W_2(D^0, D^*)$ is determined by the character $D^0$: Based on the definition of Wasserstein metrics, we can know that  $W_2(D^0, D^*)$ is positive correlative to the variance of $D^0$. It is evident that when a machine learning model is determined, sample variance is positively correlated with the variance of $D^0$. Thus, when sample variance is significant, the gain, which is brought by parallel, is remarkable. 

The feature variance is positively correlated with sample variance. Thus, the dataset with higher feature variance is suited to mini-batch SGD. Although the theorem \ref{batch_sgd_the} does not show the effect of the sample sparsity, yet we know that the feature variance is negatively correlated to sample sparsity. Thus, sparse datasets do not suit mini-batch SGD.

\textbf{The influence of $LS_{\mathcal{A}}(\mathcal{D},\mathcal{S})$} In this algorithm, the sample sequence we discuss is the sequence which build by the sample batch  and we pick the sequence  which can build the maximum $LS_{\mathcal{A}}(\mathcal{D},\mathcal{S})$. For example, in algorithm \ref{Batch SGD}, batch\_size is 3 and the sequence of server received $G_\xi(x)$ is $\{G_{\xi_1}(x_1),G_{\xi_2}(x_1),G_{\xi_3}(x_1)\}$,$\{G_{\xi_4}(x_2),G_{\xi_5}(x_2),G_{\xi_6}(x_2)\}$,...,  $\{G_{\xi_{3t-2}}(x_t),G_{\xi_{3t-1}}{\xi_2}(x_t),G_{\xi_{3t}}(x_t)\}$, where the sample or gradient in $\{\cdot,\cdot,\cdot\}$ is in on batch. Then, the $LS_{\mathcal{A}}(\mathcal{D},\mathcal{S})$ for mini-batch SGD algorithm is the $LS_{\mathcal{A}}(\mathcal{D},\mathcal{S})$ for $\xi_i,\xi_{i+1},\xi_{i+2}$ and $\xi_i,\xi_{i+1},\xi_{i+2}$ can build  a sequence whose $LS_{\mathcal{A}}(\mathcal{D},\mathcal{S})$ is the maximum in all batches.

$LS_{\mathcal{A}}(\mathcal{D},\mathcal{S})$ is small, which means that, at every iteration, most features do not gain more information from a batch, i.e., mini-batch SGD is invalid at the most feature in every iteration. Above fact suggest that when $LS_{\mathcal{A}}(\mathcal{D},\mathcal{S})$ is small, the parallel effect is poor.

\textbf{The upper bound of scalability} As we can see from theorem \ref{Batch SGD}, the gain at $t$-the iteration offered by parallel is $\frac{1}{(batch\_size)^t}$, which means that the gain growth is decreasing with the increasing of batch\_size. Although, in theory, enlarging batch\_size always gains more profit, yet the gain growths are small when batch\_size is large enough.  When the gains cannot cover the parallel cost, the scalability reaches its upper bound. The upper bound of mini-batch SGD scalability suits the first situation in the "The Upper Bound of Algorithm" section.

\subsubsection{ECD-PSGD}
To present the convergence analysis of algorithm \ref{ECD-PSGD}, we have to rewrite the goal function in to following form.
\begin{align}
f(x) &= \frac{1}{n}\sum_{i=1}^{n} F(x;\xi_i) \notag\\
&= \frac{1}{m}\sum_{j=1}^{m} \frac{1}{n_{local}} \sum_{i=0}^{n_{local}}F(x;\xi_{i,j})
\label{Problem Setting ECD and DADM}
\end{align}
And we also define following notes:
\begin{equation}
f_i(x):=\frac{1}{n_{local}}\sum_{i=0}^{n_{local}}F(x;\xi_{i,j})
\label{Sub Problem Setting ECD and DADM}
\end{equation}
\begin{equation*}
\overline{x} = \frac{1}{n}\sum_{i=1}^{n}x^(i)
\end{equation*}
\begin{equation*}
\hat{\sigma}^2 \geq \frac{1}{n_{local}}\sum_{j=1}^{n_{local}}\left\| \nabla F(x;\xi_j) - \nabla f_i(x)\right\|^2, \forall x
\end{equation*} 
\begin{equation*}
\zeta^2 \geq \frac{1}{m}\sum_{i=1}^{m}\left\| \nabla f_i(x) -f(x) \right\|^2, \forall x
\end{equation*}
\begin{align*}
\mathbb{E} \left(C(z^{(i)}_t) - z^{(i)}_t\right) = 0 ,\forall x, \forall t, \forall i\\
\widetilde{\sigma} ^2 \geq 2 \mathbb{E} \left\| C(z^{(i)}_t) - z^{(i)}_t \right\|^2 ,\forall x, \forall t, \forall i
\end{align*}
For algorithm \ref{ECD-PSGD}, Hanlin T et al.\cite{tang2018decentralization} gives following convergence theorem.
\begin{theorem}
	In algorithm \ref{ECD-PSGD}, choosing an appropriate $\gamma$, it admits
	\begin{equation}
	\frac{1}{T}\sum_{t=1}^{T} \mathbb{E}\left\| \nabla f(\overline{x}) \right\| \leq \frac{\hat{\sigma}}{\sqrt{mT}} + \frac{\hat{\sigma}\widetilde{\sigma}^2logT}{m\sqrt{mT}}+\frac{\zeta^{2/3}\widetilde{\sigma}^2logT}{mT\sqrt{T}}+h_4(\widetilde{\sigma},\zeta,T)%\frac{\widetilde{\sigma}^2logT}{T} + \frac{1}{T} +\frac{\zeta^{2/3}}{T^{2/3}} 
	\end{equation}
	\label{ECD}
\end{theorem}

As we can see from algorithm \ref{ECD-PSGD}, ECP-PSGD can be treated as the variant of mini-batch SGD: When the network $W$ is fully connected, $x = C(x), t\rightarrow \inf$,  ECD-PSGD degenerates into mini-batch SGD. Thus, ECD-PSGD inherits the character of mini-batch SGD.

\textbf{Sparsity and Feature variance} 
Following mini-batch SGD, ECD-PSGD is apt to accelerate the dataset whose variance is large ( and the dataset is dense). What is more, the $m$ is also related to $\widetilde{\sigma}$, which means the ECD-PSGD is apt to accelerate the dataset, which would lose a lot accurate during compress process.

\textbf{The influence of $LS_{\mathcal{A}}(\mathcal{D},\mathcal{S})$} The influence of similarity is the same with mini-batch SGD. 

\textbf{The upper bound of scalability} 
Again, the upper bound of scalability for ECD-PSGD shares the same characters with mini-batch SGD. As the mini-batch SGD, the profit offered by parallel is $1/\sqrt{m}$, which means that the gain growth is decreasing with the increasing of $m$. Although, in theory, enlarging m always gains more profit, yet the gain growths are small when m is large enough.  When the gains cannot cover the parallel cost, the scalability reaches its upper bound. The upper bound of ECD-PSGD scalability suits the first situation in the "The Upper Bound of Algorithm" section.

\subsubsection{DADM}
The parallel influence on the parallel stochastic gradient algorithm is reflected in the parameters in the theorem. However, DADM uses different proof structure to offer the convergence conclusion. In the proof of DADM: different workers solve a local problem, i.e., $f_i(x)$ in Eq. \ref{Sub Problem Setting ECD and DADM} at each iteration and then broadcast its information to other workers to solve globe problem $f(x)$ in Eq. \ref{Problem Setting ECD and DADM}. What is more, DADM is to find the expected duality gap. Thus, the convergence analysis conclusion is unrelated to a dataset and machine learning model character, and the parallel influence is buried in the problem setting instead of directly convergence theorem.  The convergence theorem about DADM is the conclusion from the work\cite{Zheng2017A}. 

\begin{theorem}
	$f(\cdot)$ , $\xi_i$ and $\Delta \alpha_{local}$ satisfy some requirements. When  $t$ satisfies following condition, the expected duality gap of goal function and its dual form is smaller than $\epsilon$
	\begin{align*}
	t&\geq (h_5(\xi_i,\gamma,\lambda)+\frac{1}{local\_batch\_size*m})log\bigg((h_5(\xi_i,\gamma,\lambda)\notag \\
	&+\frac{1}{local\_batch\_size*m})h_6(x_0,\epsilon)\bigg)
	\end{align*}
	\label {DADM the}
\end{theorem}

\textbf{Sample Diversity}  
As we can see from the proof, the primary purpose of parallel technology is to cut the original problem into several subproblems. Thus, from the aspect of subproblem, the parallel algorithm will fail to accelerate the algorithm when some nodes solve the same problem. To ensure different nodes solve different subproblems, we should ensure the dataset is high sample diversity. For example, when a dataset consists of little kinds of the sample, i.e., the dataset is the replication of a little sample, the sub-dataset in each node in the cluster would be almost the same, which means $f_i(x), \forall i$ in eq. \ref{Sub Problem Setting ECD and DADM} are the same. In this case, DADM  fails to make full use of multi-nodes. Thus, we can know that DADM is apt to accelerate the dataset whose sample diversity is high.

\textbf{The influence of $LS_{\mathcal{A}}(\mathcal{D},\mathcal{S})$} The influence of similarity is hard to be shown in theory analysis. However, from algorithm \ref{DADM} description step 2 in SERVER part, we can observe that using the definition of $LS_{\mathcal{A}}(\mathcal{D},\mathcal{S})$ in mini-batch SGD, when $LS_{\mathcal{A}}(\mathcal{D},\mathcal{S})$ is small,  $v^t_{local}$s from a different worker would be almost the same, which would decrease the influence of parallel.   Above fact suggest that when $LS_{\mathcal{A}}(\mathcal{D},\mathcal{S})$ is small, the parallel effect is poor.

\textbf{The upper bound of scalability} 
Again, the upper bound of scalability for DADM shares the same characters with mini-batch SGD. As the mini-batch SGD, the profit offered by parallel is $1/m$, which means that the gain growth is decreasing with the increasing of $m$. Although, in theory, enlarging $m$ always gains more profit, yet the gain growths are small when $m$ is large enough.  When the gains cannot cover the parallel cost, the scalability reaches its upper bound. The upper bound of DADM scalability suits the first situation in the "The Upper Bound of Algorithm" section.

\subsection{Theorem Proof}
\textbf{Theorem 1}
The minimum of the maximum of $\tau$ is the number of workers, i.e., $m \leq \tau_{max}$\

\textbf{proof}
In a $M$ worker cluster, for the slowest worker, at $t$th iteration, this slowest worker submits its gradient to the server, At this time, other workers are in computing their gradient. At $t+j$th iteration, the slowest workers submit its gradient again. At this time, other workers are already submitted at least one gradient in $j$ iterations, i.e., $j > M$. Thus, an asynchronous parallel system at least has a $M$ iteration delay. And when all workers share the same performance, the system would achieve the minimum.

\textbf{Theorem 3}
When goal function Eq. \ref{reg_goal} is running on algorithm \ref{Batch SGD}, then we have
\begin{align}
&\mathbb{E}_{x_t\in D_t}f(x_t) - f(x^*) \leq \notag\\
&\bigg((h_2\left(F\left(\cdot\right)\right)\left(d(\mu_{D^t},\mu_{D^*})+\frac{\sigma_{D^*}+W_2(D^0,D^*)(1-\gamma\lambda)^t}{(batch\_size)^{t/2}}\right)\\
&+h_3\left(F\left(\cdot\right)\right)\bigg)^2
\end{align}
where $D^t$ is the distribution of $x^t$, $D^*$ is the distribution of $x^*$, $sigma_{D}$ is the standard deviation of distribution $D$, $\mu_D$ is the mean of distribution $D$. $W_2(D_1,D_2)$ is the wasserstein metrics between $D_1$ and $D_2$

\begin{proof}
	Based on the work by M.Zinkevich et al\cite{Zinkevich2010Parallelized}, we treat $x_t$ as random variable firstly and its distribution is $D_t$. We have following theorem (Theorem 11 in M.Zinkevich \cite{Zinkevich2010Parallelized})
	Given a cost function $f$ such that $\left\|f \right\|_L$ and$\left\|\nabla f \right\|_L$( $\left\| \cdot\right\|_L$  is Lipschitz seminorm ) are bounded, a distribution $D$ such that $\sigma_D$ and is bounded , then ,for any $v$
	\begin{align}
	&\mathbb{E}_{x\in D} [f(x)] - \mathop{min}_x f(x) \leq \notag\\ &(W_2(v,D))\sqrt{2\left\|\nabla f \right\|_L(f(v)-\mathop{min}_x f(x))} \notag\\
	&+\left\|\nabla f \right\|_L(W_2(v,D))^2/2 + (f(v)-\mathop{min}_x f(x))
	\end{align}
	
	When $v=\mu_{D^*}$ $W_2(\mu_{D^*},D)$ is the  relative standard deviation of $x_t$ with respect to $\mu_{D^*}$, i.e. $\sigma_D^{\mu_{D^*}}$.
	
	Based on Theorem 32 in M.Zinkevich et al\cite{Zinkevich2010Parallelized}, we know that 
	\begin{equation}
	\sigma_D^{\mu_{D^*}}\leq \sigma_D + d(\mu_{D^*},\mu_{D}) 
	\end{equation}
	
	\begin{equation}
	\sigma_{D^t} \leq \sigma_{D^*} + W(D^*,D^0)(1-\lambda\gamma)^t 
	\end{equation}
	
	Suppose that random variable $X^1,X^2,X^3,...,X^k$
	are independent and identically distributed. if $A=\frac{1}{k}\sum_{i=1}^{k}X^i$, it is the case that:
	\begin{align*}
	\mu_A=\mu_{X^1}=\mu_{X^2}=...=\mu_{X^k}\\
	\sigma_A\leq\frac{\sigma_{X^1}}{\sqrt{k}}
	\end{align*}

	As we can see from $x_t$, before average operation, $x^{i}_t$ is  independent and identically distributed random variable. In each iteration, $\sigma_{D^t}$ is shrinked $\frac{1}{batch\_size}$. Combining above equations, we can get theorem.
\end{proof}

\textbf{Lemma } $LS_{\mathcal{A}}(\mathcal{D},\mathcal{S})$ is positively correlated to the scalability in Hogwild!
\begin{proof}
	In the Hogwild! proof, the $\tau$ is created in the following equation, figure \ref{Hogwild!_proof}. In following equation, $\delta, \Omega, \rho$ is create by the sum of multiplication of gradient $G_{\xi_i}$ or model difference ($x_i$ - $x_{k(i)}$, which can be described as  $G_{\xi_i}$). The sum range is $\xi_i$ to $\xi_{i--\tau}$. The above facts show that the original definition $\delta, \Omega, \rho$  is large: it is unnecessary to calculate those parameters in the whole dataset. just it is better define those parameter server in sample sequence neighborhood $\tau_{max}$ samples sub-dataset: If we define  $\delta, \Omega, \rho$  as $\delta_{}, \Omega_{local}, \rho_{local}$, which is calculated in  sample sequence neighborhood $\tau_{max}$ samples sub-dataset and replace  $\delta, \Omega, \rho$ in Hogwild@ proof, the whole proof of Hogwild! is still sound. So, we find a tighter upper bound of Hogwild! Algorithm.
	
	\begin{figure}[!htbp] 
		\centering 
		\subfigure [A8 in Hogwild! proof]{   
			\includegraphics[width=1\columnwidth]{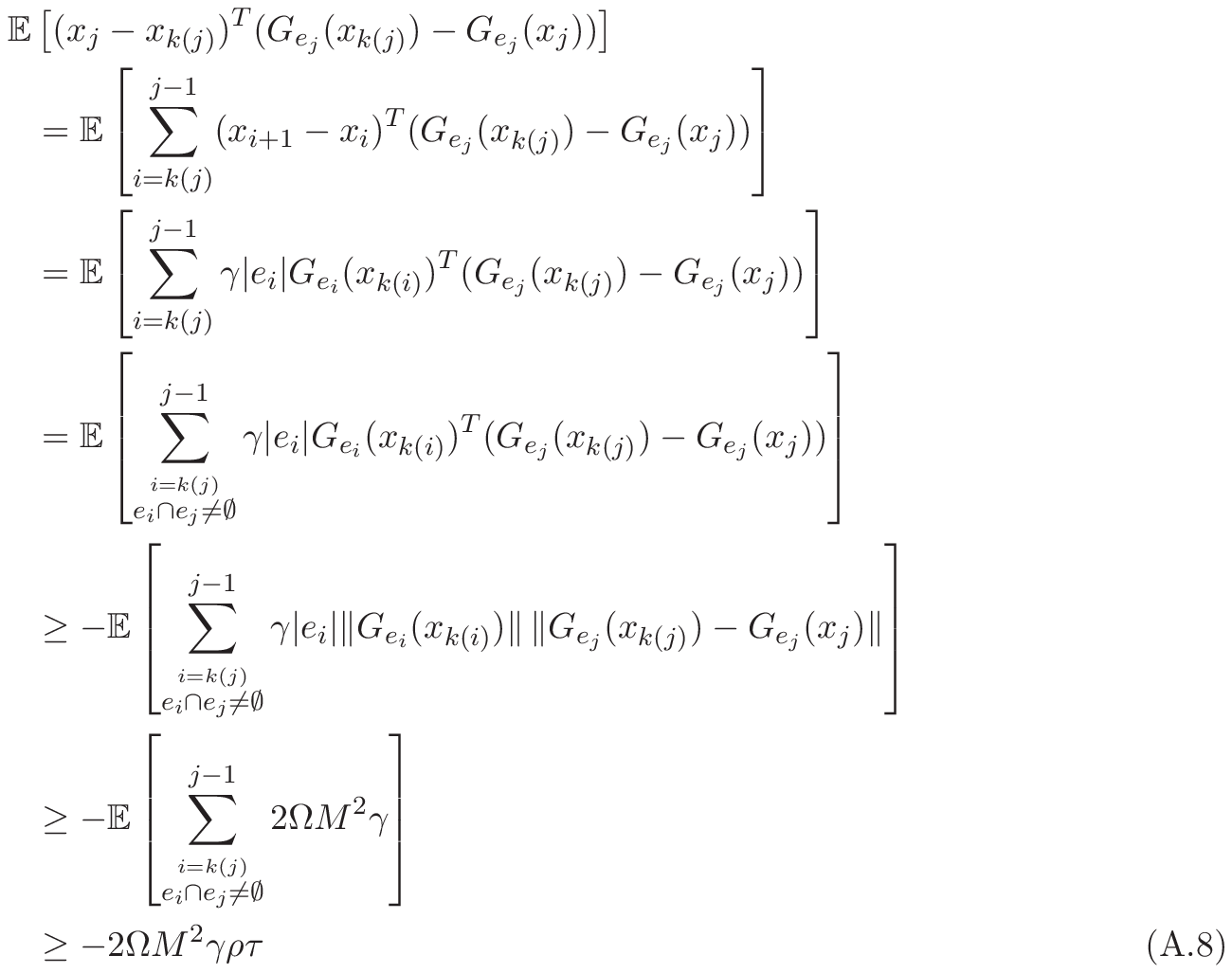} 
		} 
		\subfigure [part of Hogwild! proof]{ 
			\includegraphics[width=1\columnwidth]{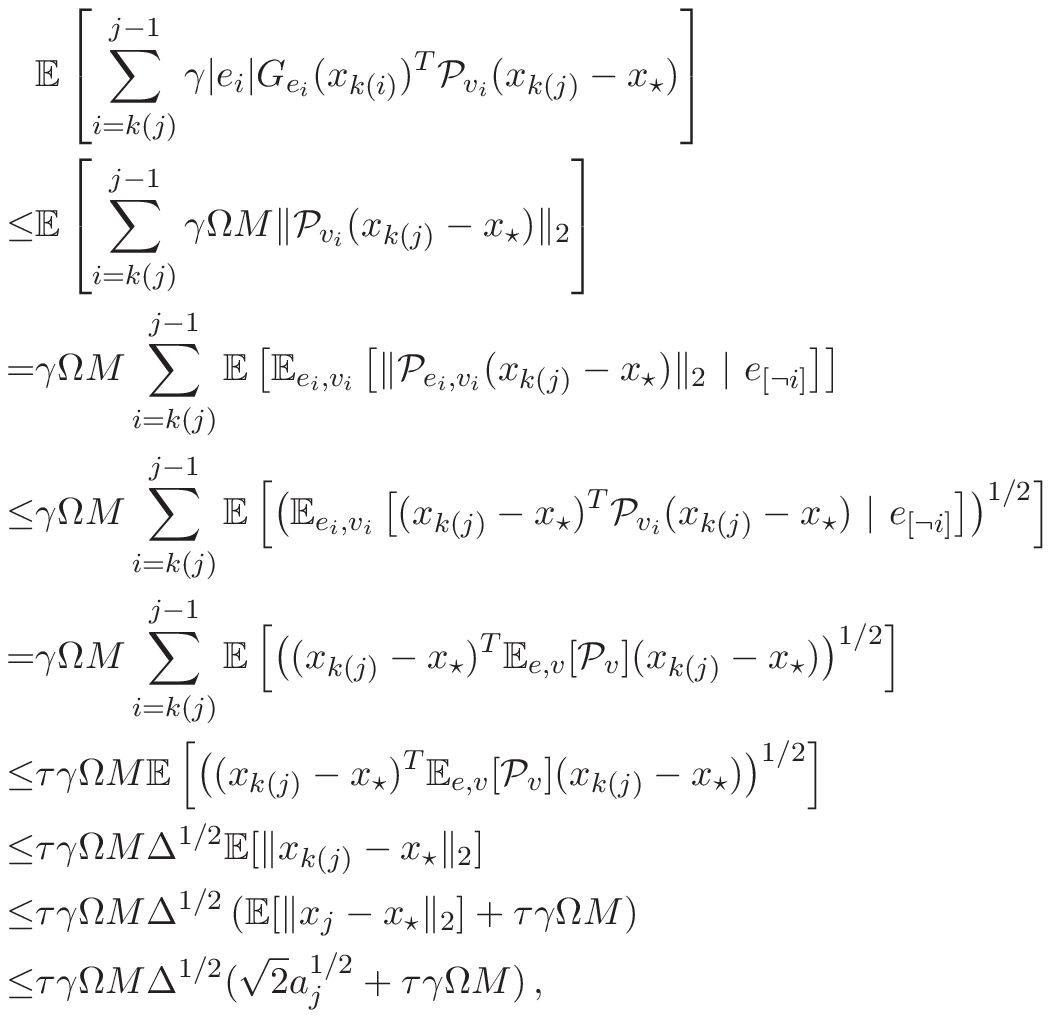} }
		\subfigure [A6 in Hogwild! proof] { 
			\includegraphics[width=1\columnwidth]{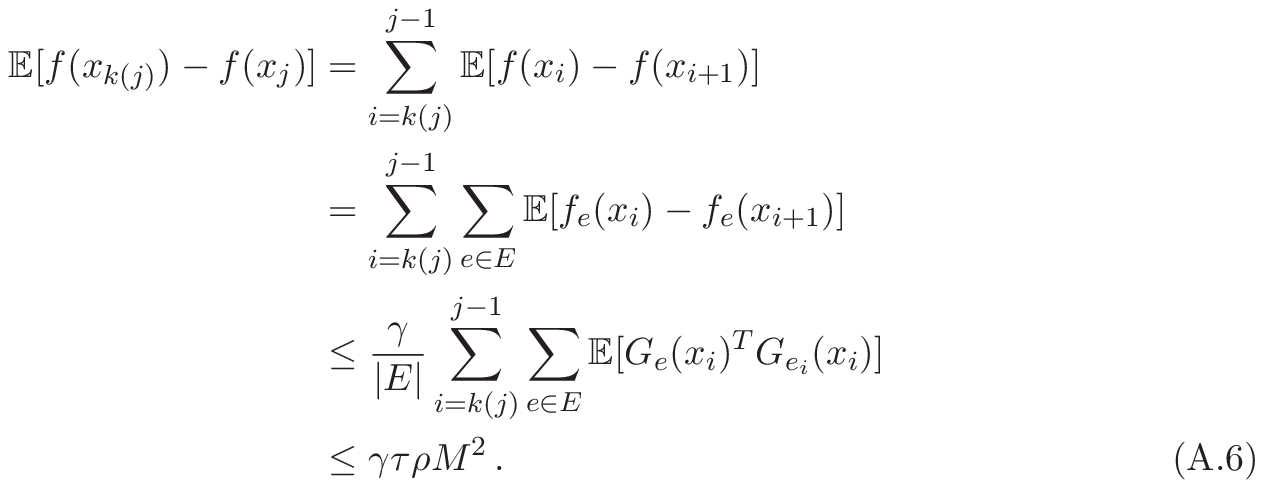} 
		} 
		\caption{ the proof segment of Hogwild! } 
		\label{Hogwild!_proof} 
	\end{figure}

	As we can see from the definition, when $LS_{\mathcal{A}}(\mathcal{D},\mathcal{S})$  is small,  $\delta_{local}, \Omega_{local}, \rho_{local}$ is also small, which would increase the scalability ability.
	
\end{proof}

\end{document}